\documentclass[fleqn,usenatbib]{mnras}

\usepackage{newtxtext,newtxmath}

\usepackage[T1]{fontenc}

\DeclareRobustCommand{\VAN}[3]{#2}
\let\VANthebibliography\thebibliography
\def\thebibliography{\DeclareRobustCommand{\VAN}[3]{##3}\VANthebibliography}

\usepackage{float}
\usepackage{epsfig}
\usepackage{graphicx}
\usepackage{graphics}
\usepackage{latexsym}
\usepackage{hyperref}
\usepackage{wasysym}
\usepackage{threeparttable}
\usepackage{multirow}

\def\gsim{\;\lower4pt\hbox{${\buildrel\displaystyle >\over\sim}$}\;}
\def\lsim{\;\lower4pt\hbox{${\buildrel\displaystyle <\over\sim}$}\;}
\def\grls{\;\lower4pt\hbox{${\buildrel\displaystyle >\over <}$}\;}

\title[SN 2022vqz]{SN 2022vqz: A Peculiar Subluminous Type Ia Supernova with Prominent Early Excess Emission}

\author[G. Xi et al.]{
Gaobo Xi,$^{1}$
Xiaofeng Wang,$^{1,2}$\thanks{E-mail: wang\_xf@mail.tsinghua.edu.cn}
Gaici Li,$^{1}$
Jialian Liu,$^{1}$
Shengyu Yan,$^{1}$
Weili Lin,$^{1}$
Jieming Zhao,$^{1}$
\newauthor
Alexei V. Filippenko,$^{3}$
WeiKang Zheng,$^{3}$
Thomas G. Brink,$^{3}$
Y. Yang,$^{3}$
Shuhrat A. Ehgamberdiev,$^{4,5}$
\newauthor
Davron Mirzaqulov,$^{4}$
Andrea Reguitti,$^{6,7}$
Andrea Pastorello,$^{7}$
Lina Tomasella,$^{7}$
\newauthor
Yongzhi Cai,$^{8,9,10}$
Jujia Zhang,$^{8,9,10}$
Zhitong Li,$^{11,12}$
Tianmeng Zhang,$^{11}$
\newauthor
Hanna Sai,$^{1}$
Zhihao Chen,$^{1}$
Qichun Liu,$^{1}$
Xiaoran Ma,$^{1}$
and Danfeng Xiang$^{1}$
\\
$^{1}$Physics Department and Tsinghua Center for Astrophysics, Tsinghua University, Beijing 100084, China\\
$^{2}$Beijing Planetarium, Beijing Academy of Science and Technology, Beijing 100044, China\\
$^{3}$Department of Astronomy, University of California, Berkeley, CA 94720-3411, USA\\
$^{4}$Ulugh Beg Astronomical Institute, Uzbekistan Academy of Sciences, Tashkent 100052, Uzbekistan\\
$^{5}$National University of Uzbekistan, Tashkent 100174, Uzbekistan\\
$^{6}$INAF - Osservatorio Astronomico di Brera, Via E. Bianchi 46, 23807 Merate (LC), Italy\\
$^{7}$INAF - Osservatorio Astronomico di Padova, Vicolo dell'Osservatorio 5, 35122 Padova, Italy\\
$^{8}$Yunnan Observatories, Chinese Academy of Sciences, Kunming 650216, China\\
$^{9}$Key Laboratory for the Structure and Evolution of Celestial Objects, Chinese Academy of Sciences, Kunming 650216, China\\
$^{10}$International Centre of Supernovae, Yunnan Key Laboratory, Kunming 650216, China\\
$^{11}$Key Laboratory of Optical Astronomy, National Astronomical Observatories, Chinese Academy of Sciences, Beijing 100101, China\\
$^{12}$School of Astronomy and Space Science, University of Chinese Academy of Sciences, Beijing 101408, China\\
}

\date{Accepted XXX. Received YYY; in original form ZZZ}

\pubyear{2023}

\begin{document}
\label{firstpage}
\pagerange{\pageref{firstpage}--\pageref{lastpage}}
\maketitle

\begin{abstract}

We present extensive photometric and spectroscopic observations of the peculiar Type Ia supernova (SN~Ia) 2022vqz. It shares many similarities with the SN 2002es-like Type Ia supernovae (SNe~Ia), such as low luminosity ($M_{B,\rm max}=-18.11\pm0.16$ mag) and moderate post-peak decline rate ($\Delta m_{15,B}=1.33\pm0.11$ mag). The nickel mass synthesised in the explosion is estimated as $0.20\pm0.04~{\rm M}_\odot$ from the bolometric light curve, which is obviously lower than that of normal SNe~Ia. SN~2022vqz is also characterised by slowly expanding ejecta, with Si~II velocities persisting around 7000~km~s$^{-1}$ since 16 days before peak brightness, unique among all known SNe~Ia. While all of these properties imply a lower-energy thermonuclear explosion that should leave a considerable amount of unburnt materials, the absent signature of unburnt carbon in spectra of SN~2022vqz is puzzling. A prominent early peak is clearly detected in the ATLAS $c$- and $o$-band light curves and in the ZTF $gr$-band data within days after the explosion. Possible mechanisms for the early peak are discussed, including the sub-Chandrasekhar-mass double-detonation model and interaction of SN ejecta with circumstellar material. We find that both models face some difficulties in replicating all aspects of the observed data. As an alternative, we propose a hybrid C-O-Ne white dwarf as the progenitor of SN 2022vqz; it   can simultaneously reconcile the tension between low ejecta velocity and the absence of carbon. We further discuss the diversity of SN 2002es-like objects and their origin in the context of different scenarios.
\end{abstract}




\section{Introduction} \label{sec:intro}

Although it is commonly accepted that Type Ia supernovae (SNe~Ia) are thermonuclear explosions of carbon-oxygen (C-O) white dwarfs (WDs) in close binary systems \citep{1960ApJ...132..565H,2012ApJ...744L..17B}, the detailed properties of their progenitors, as well as the exact explosion mechanism, are still inconclusive \citep{2018PhR...736....1L,2023RAA....23h2001L}. Observationally, about 70\% of SNe~Ia display relatively homogeneous properties, especially after some empirical corrections based on their light-curve decline rates \citep{1993ApJ...413L.105P} and colours \citep{2007A&A...466...11G,2007ApJ...659..122J,2005ApJ...620L..87W}. These SNe~Ia are known as ``Branch-normal'' ones \citep{1993AJ....106.2383B} and are used as excellent cosmological standardisable candles \citep{1998AJ....116.1009R,2022ApJ...934L...7R,1998ApJ...507...46S,1999ApJ...517..565P}. Note that even Branch-normal SNe~Ia can be further classified into several subclasses that may also have different physical origins \citep{2005ApJ...623.1011B,2009ApJ...699L.139W,2013Sci...340..170W,2019ApJ...882..120W}.

Beside the normal ones, the remaining SNe~Ia show a substantial  diversity, and are classified into different subclasses depending on their photometric and spectroscopic properties \citep{1997ARA&A..35..309F,2014Ap&SS.351....1P,2017hsn..book..317T}. In particular, the subluminous group represented by SN~1991bg  (``91bg-like") are characterised by their rapid evolving light curves, low ejecta velocity, more prominent intermediate-mass elements (IMEs), and strong Ti~II absorption in spectra at maximum light \citep{1992AJ....104.1543F,1993AJ....105..301L}. These peculiarities are generally interpreted as signatures of a lower burning efficiency and a cooler ejecta temperature \citep{1997MNRAS.284..151M,2009MNRAS.399.1238H}.

Another peculiar subluminous subclass of SNe~Ia, which are represented by SN~2002es (dubbed ``02es-like"), share most of the characteristics with the 91bg-like objects, except for displaying broader light curves that are comparable to those of normal SNe~Ia \citep{2012ApJ...751..142G,2015ApJ...799...52W}. The normal light-curve decline rate combined with the low peak luminosity put 02es-like objects at a position that was long believed to be devoid of events on the Phillips-relation diagram \citep{1993ApJ...413L.105P}.
In comparison with other subtypes, the subclass of 02es-like SNe~Ia is relatively rare, contributing to $\sim 2.5$\% of all SNe~Ia \citep{2012ApJ...751..142G}. By now, only about a dozen  events have been identified as 02es-like \citep{2021ApJ...919..142B}. The prototype SN~2002es itself shows a subluminous peak luminosity ($M_{B} \approx -17.78$ mag) and a slow ejecta velocity ($v_{\rm ej} \approx$ 6000 km~s$^{-1}$ around maximum light). Its optical spectra are also characterised by strong Si~II $\lambda$5972, O~I, and Ti~II lines, similar to those of the 91bg-like SNe~Ia. It is expected that the 02es-like SNe~Ia should have subluminous peaks and slowly-expanding ejecta. However, some events with properties close to those of normal ones were also classified as 02es-like. For example, the peak magnitudes of SN~2006bt \citep{2017hsn..book..317T}, SN~2016jhr \citep{2017Natur.550...80J}, and SN~2022ywc \citep{2023arXiv230806019S} are as bright as $M_{B}\approx -18.8$ mag. The Si~II velocities of SN~2006bt \citep{2010ApJ...708.1748F} and PTF10ops \citep{2011MNRAS.418..747M} measured near maximum light are $\sim 10,000$ km~s$^{-1}$. 
Another proposed candidate is SN~2019yvq \citep{2020ApJ...898...56M,2021ApJ...914...50T,2021ApJ...919..142B}, which showed an unprecedented high velocity of $\sim 15,000$ km~s$^{-1}$. This greatly extends the range of 02es-like SNe~Ia if they have similar origins. All of the above-mentioned events show normal light curves but have the spectral characteristics of low-luminosity SNe~Ia, especially the distinct Ti~II absorption. These commonalities make them classified as 02es-like SNe~Ia and demonstrate the diversity in this subclass.

Observations of 02es-like SNe~Ia revealed many other properties of this subclass. The nebular spectra of 02es-like SN~2010lp \citep{2013ApJ...775L..43T} and iPTF14atg \citep{2016MNRAS.459.4428K} are characterised by broad and strong [Fe~II] and [Ca~II] emission, as in low-luminosity SNe~Ia. However, in contrast to the 91bg-like events, they do not display narrow [Fe~III] or [Co~III] emission but rather narrow [O~I] in their nebular spectra. The light curves of SN~2002es show a rapid decline from $\sim 1$ month after the peak \citep{2012ApJ...751..142G}, which is not observed in other 02es-like SNe~Ia. \cite{2015ApJ...799...52W} noticed that the 02es-like SNe~Ia tend to reside in luminous, early-type galaxies with minimal star formation, and located far from the centre of their hosts, suggesting longevity of their progenitor systems \citep{2017hsn..book..317T}. There are also outliers, such as SN~2016ije, located in a bluer, low-mass, star-forming galaxy \citep{2023ApJ...950...17L}.

More recent studies indicate that all the 02es-like SNe~Ia with good early photometric coverage, including iPTF14atg \citep{2015Natur.521..328C}, SN~2016jhr, 2019yvq, and 2022ywc, show early blue/ultraviolet (UV) excesses/peaks within days after the explosion. An early-time excess has been considered to be evidence for a nondegenerate companion in the single-degenerate (SD) scenario \citep{1973ApJ...186.1007W}, interacting with the expanding ejecta and creating additional UV/optical flux. However, owing to the effect of the viewing-angle distribution, the excess is predicted to be observable in only $\sim 10$\% of the events \citep{2010ApJ...708.1025K}. This prediction is more consistent with the observed rate of early excesses in normal SNe~Ia \citep{2022arXiv220707681B}, which is $\sim 20$\% \citep{2022MNRAS.512.1317D,2022MNRAS.513.3035M}.

A promising scenario for 02es-like SNe~Ia is that they arise from a double-degenerate (DD) system, in which the companion of the exploding WD is another He-C-O WD \citep{1984ApJS...54..335I,1984ApJ...277..355W}. \cite{2011MNRAS.418..747M} and \cite{2012ApJ...751..142G} suggested that PTF10ops and SN~2002es can be explained by a violent-merger model of two 0.9~${\rm M}_\odot$ WDs \citep{2010Natur.463...61P}. \cite{2013ApJ...778L..18K,2016MNRAS.459.4428K} found that a revised merger model with $0.9+0.76~{\rm M}_\odot$ WDs and different metallicities provide both light curves and spectra that match the observations of SN~2010lp and iPTF14atg. This model also leaves unburnt O near the centre of the ejecta, potentially reproducing the [O~I] emission in the nebular spectra of 02es-like events. However, the merger model does not naturally explain the universal appearance of early-time excesses. A possible solution refers to the interaction between ejecta and extended circumstellar material (CSM; \citealt{2016ApJ...826...96P,2021ApJ...909..209P}). The CSM may be ejected before the merging state as tidal tails, disc winds, or jets \citep{2013ApJ...772....1R,2015MNRAS.447.2803L}. 

The DD scenario has a quite large parameter space. If the two merging WDs have larger masses, they can potentially result in a more-luminous super-$M_{\rm Ch}$ event, falling into the category of SN 2003fg-like SNe~Ia \citep{2006Natur.443..308H,2010ApJ...722L.157V,2010ApJ...713.1073S}. There are also some similarities between 02es-like and 03fg-like subtypes that are distinct from other SNe~Ia, such as nonmonotonic early-time bumps and unique UV colour evolution, suggesting that they may originate from a single DD scenario with different WD masses \citep{2023arXiv230911563H}.

An alternative explanation proposed for the early peak of 02es-like objects is the double-detonation model, in which a sub-Chandrasekhar-mass (sub-$M_{\rm Ch}$) C-O WD is detonated by ignition of a surface He layer \citep{2019ApJ...873...84P}. A surface He layer of $\sim 0.05~\rm{M}_\odot$ was adopted to explain the early-time excesses of SN~2016jhr \citep{2017Natur.550...80J} and SN~2019yvq \citep{2020ApJ...898...56M,2020ApJ...900L..27S,2021ApJ...919..142B}. However, in the case of SN~2019yvq, the double-detonation model faces some issues that are difficult to reconcile \citep{2021ApJ...914...50T}. The inferred WD masses from early-time photometry (0.96 M$_\odot$; \citealt{2020ApJ...898...56M}) and nebular-phase spectroscopy (1.15~M$_\odot$; \citealt{2020ApJ...900L..27S}) show a large discrepancy. Observing along the detonation pole of a thick He layer may reconcile the overluminous-model prediction of a more-massive WD \citep{2020ApJ...900L..27S}, but this will also introduce additional peculiar signals that were not observed in SN~2019yvq, such as severe line blanketing from He ashes \citep{2021ApJ...914...50T}.

In this paper, we present our observations and analysis of SN~2022vqz, a subluminous 02es-like SN~Ia with a remarkable early peak and extremely low ejecta velocity. Section~\ref{sec:obs} describes our observations and data reduction. Detailed analysis of the photometry and spectroscopy is presented in Section~\ref{sec:photo} and Section~\ref{sec:spec}, respectively. In Section~\ref{sec:dis}, we discuss possible explosion models for SN~2022vqz. We summarise the main results in Section~\ref{sec:sum}.

\section{Observations and Data Reduction} \label{sec:obs}

\begin{figure*}
\center
\includegraphics[angle=0,width=.8\textwidth]{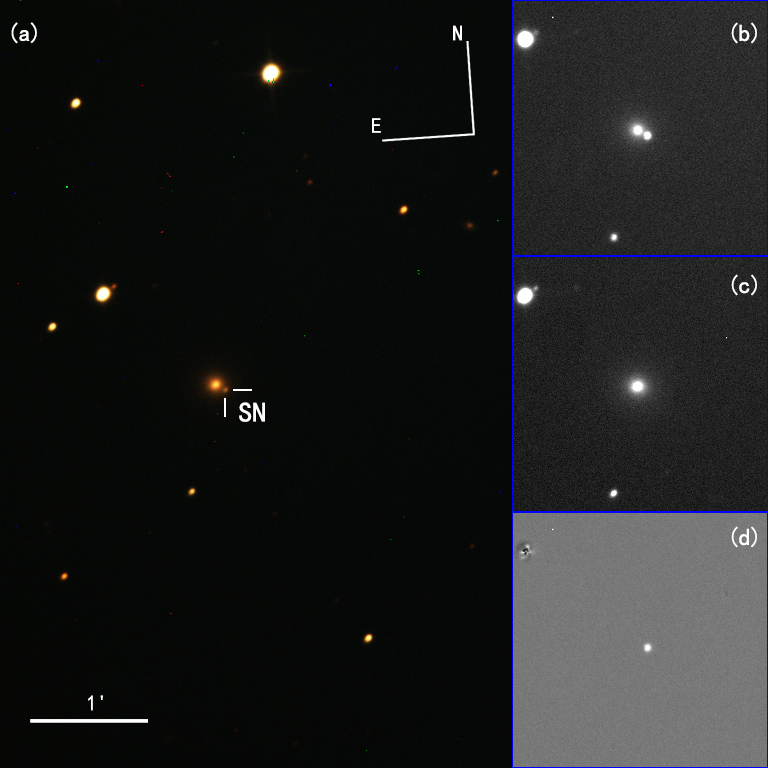}
\vspace{0.0cm}
\caption{(a) A $BVr$-band colour image of SN~2022vqz (marked with the white crosshair) with its host galaxy MCG+05-03-011, taken with the TNT 80~cm telescope on MJD 59849.80 ($\sim 14$~days before $B$ maximum). A scale bar is shown in the bottom-left corner. North is up and east is to the left.
(b) $V$-band image on MJD 59876.68. (c) $V$-band template constructed from the MJD 59849.80 image. (d) Result of the template subtraction.}
\label{figVimg} \vspace{-0.0cm}
\end{figure*}

SN~2022vqz was discovered at 18.38 mag on MJD 59846.369 (2022 Sep. 24.369; UTC dates are used throughout this paper) by the Zwicky Transient Facility (ZTF; \citealt{2022TNSTR2766....1F})  in the $g$ band. This transient was also detected by the Asteroid Terrestrial-impact Last Alert System \citep[ATLAS; ][]{2018PASP..130f4505T,2020PASP..132h5002S} survey on MJD 59845.471, $\sim 1$ day earlier than the ZTF discovery, at $\sim 19$~mag in their $o$ band. The last nondetection from the ATLAS survey was on MJD 59844.484, posting a strict constraint on its explosion time. Later, two spectra taken by \cite{2022TNSCR2778....1T} and \cite{2022TNSCR2845....1M} classified this transient as a peculiar SN~Ia. It is located at J2000 coordinates $\alpha=00^{\rm h}51^{\rm m}05^{\rm s}.78$, $\delta=+29^\circ 32'12''.2$, which is $4''.9$ west and $3''.2$ south of the centre of its host galaxy MCG+05-03-011, a lenticular (S0) galaxy with a heliocentric redshift of $z=0.016995$. Figure~\ref{figVimg}(a) shows the $BVr$-band composite image of SN 2022vqz.

\subsection{Photometric Data} \label{sec:obsphoto}

We performed follow-up optical photometric observations of SN~2022vqz with multiple facilities, including the Tsinghua-NAOC 80~cm telescope at Xinglong Observatory of NAOC \citep[TNT; ][]{2008ApJ...675..626W,2012RAA....12.1585H}, the AZT-22 1.5~m telescope at Maidanak Observatory \citep{2018NatAs...2..349E}, and the Schmidt 67/92~cm and the 1.82~m Copernico Telescope at Cima Ekar Observing Station (Padova). Our multiband photometric monitoring began on 2022 Sep. 27.80 (MJD 59849.80) and has good coverage over a period of $\sim 2.5$ months. Standard \texttt{IRAF}\footnote{IRAF (Image Reduction and Analysis Facility) is distributed by the National Optical Astronomy Observatories (NOAO), which are operated by the Association of Universities for Research in Astronomy (AURA), Inc., under cooperative agreement with the U.S. National Science Foundation.} routines were adopted to reduce the CCD images, including bias and flat-field corrections. The magnitudes are calibrated using a set of nearby reference stars from the Sloan Digital Sky Survey (SDSS) catalogue \citep{2000AJ....120.1579Y,2006AJ....131.2332G}. The number of reference stars ranges from 30 to 40 for the different facilities. For flux calibration of the $UBVRI$-band photometry from AZT and $BV$-band photometry from TNT, the SDSS $ugriz$ magnitudes of the reference stars are converted to  standard $UBVRI$ magnitudes\footnote{http://classic.sdss.org/dr4/algorithms/sdssUBVRITransform.html\#Lupton2005}.

Although SN~2022vqz is offset from the centre of its host galaxy, background-light contamination is still prominent, especially for its late-time light curves. Usually, a host-galaxy template can be taken when the SN fades away, but this will take a long time for SN 2022vqz. As an alternative, we construct the host-galaxy templates of SN 2022vqz using two different methods. The first is smoothly interpolating the host-galaxy flux at the location of the SN by solving a diffusive equation bounded by surrounding starlight \citep{2004AJ....128.1857Z}. The second is using the flux information from the opposite side as a template since the host galaxy looks symmetric. The subtraction process follows \texttt{Zrutyphot} (Mo et al., in prep.). With the templates constructed by the above two methods, the final results of the photometry are consistent. Figure~\ref{figVimg} shows an example of an original image, constructed template, and their difference image.

\begin{figure*}
\center
\includegraphics[angle=0,width=0.7\textwidth]{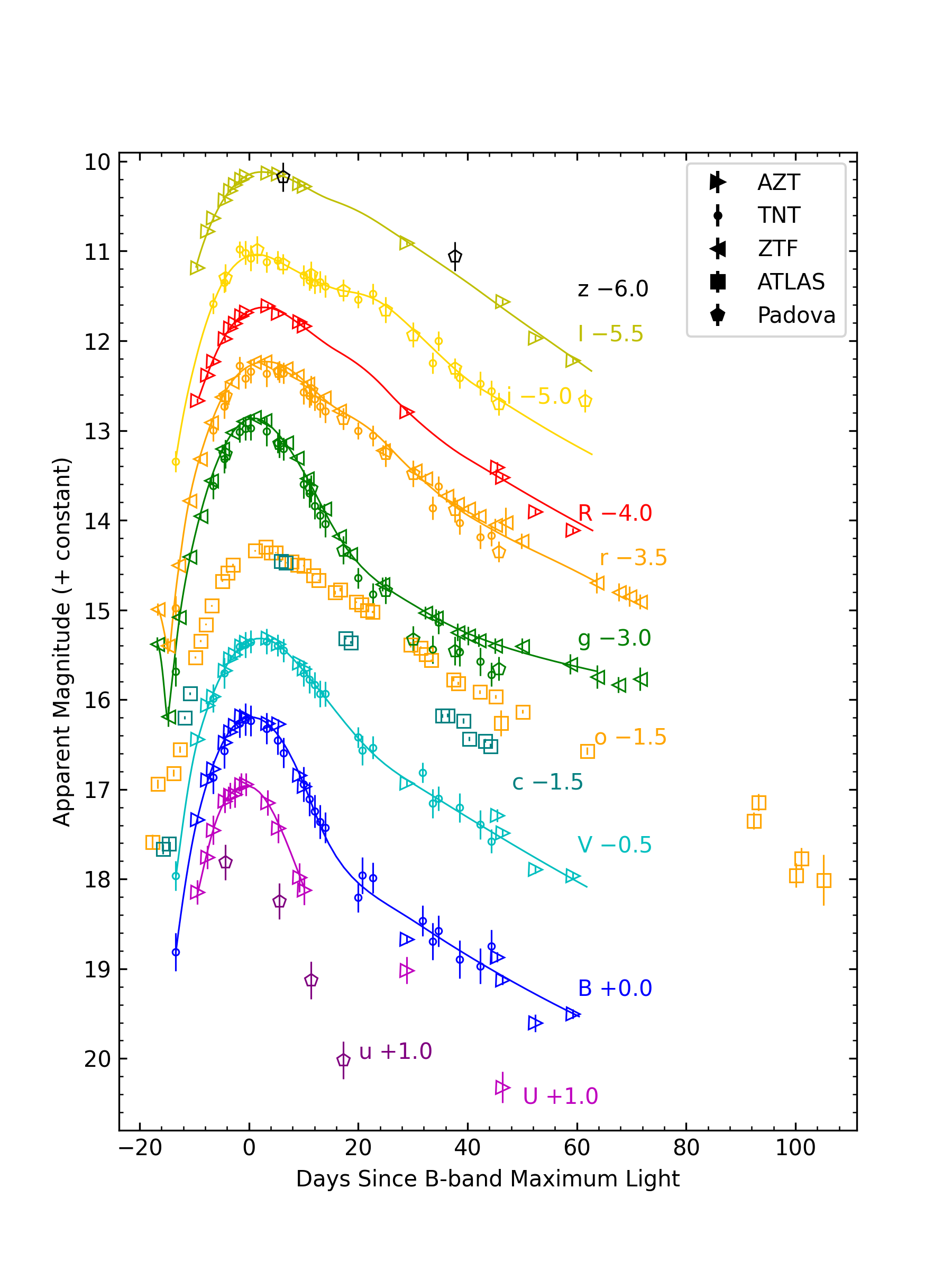}
\vspace{0.0cm}
\caption{The multiband light curves of SN~2022vqz. Data from different sources are shown with different symbols. Days are relative to the $B$-band maximum light on 2022-10-11.43 (MJD~59863.43), corrected to the rest frame by a factor $1+z=1.017$. Solid lines, which are template light curves constructed from the 02es-like SN~Ia sample in Figure~\ref{figlccmp}, and adjusted to match the light curves of SN~2022vqz, are used to guide the eye.}
\label{figlc} \vspace{-0.0cm}
\end{figure*}

In our analysis, we also included the publicly available $gr$-band forced photometry from ZTF \citep{2019PASP..131a8003M}\footnote{https://lasair-ztf.lsst.ac.uk/object/ZTF22abhrjld/} and $co$-band ({\it cyan}, {\it orange}) forced photometry from ATLAS. These photomeric results were obtained after subtraction of the corresponding template images taken before the SN explosion, and hence are more reliable. The first observation of ATLAS was on 2022 Sep. 23.471 (MJD 59845.471), less than 1 day after the latest nondetection. Combined with the early ZTF observations which started $\sim 1$ day later, the early-time photometric evolution of SN~2022vqz can be well constrained. Figure~\ref{figlc} shows the multiband light curves of SN~2022vqz.  

\subsection{Spectroscopic Data} \label{sec:obsspec}

Our spectroscopy of SN~2022vqz covered phases from $-7.1$ to $+97.4$ days relative to its peak brightness. We obtained 9 spectra with BFOSC mounted on the 2.16~m telescope of the Xinglong Observatory (XLT), 5 spectra with YFOSC mounted on the Lijiang 2.4~m telescope \citep[LJT; ][]{2015RAA....15..918F} of Yunnan Observatories (YNAO), 3 spectra with the Kast spectrograph\citep{miller1994kast} on the 3~m Shane reflector at Lick Observatory, and 1 spectrum with AFOSC mounted on the 1.82~m Copernico Telescope of Padova. In addition, 1 spectrum was taken of the host galaxy MCG+05-03-011 with XLT+BFOSC. The Lick/Kast spectra were taken with the slit oriented along the parallactic angle to minimise the effects of atmospheric dispersion \citep{1982PASP...94..715F}. The journal of spectroscopic observations is shown in Table~\ref{tabspec}.

Standard \texttt{IRAF} routines were adopted for reduction of all the optical spectra taken by LJT, XLT, Lick 3~m Shane, and Padova 1.82~m Copernico telescopes, including bias, flat-field corrections, and removal of cosmic rays.  The Copernico spectrum is reduced using a dedicated pipeline \texttt{foscgui}\footnote{\texttt{foscgui} is a graphic user interface aimed at extracting SN spectroscopy and photometry obtained with FOSC-like instruments. It was developed by E. Cappellaro. A package description can be found at http://sngroup.oapd.inaf.it/foscgui.html .}. The wavelength scale was calibrated with comparison-lamp spectra, and the fluxes were calibrated with standard stars observed on the same night at similar airmasses. All spectra were further corrected for atmospheric extinction using the extinction curves obtained at each observatory, and telluric lines were also removed from the spectra.

\begin{figure*}
\center
\includegraphics[angle=0,width=0.7\textwidth]{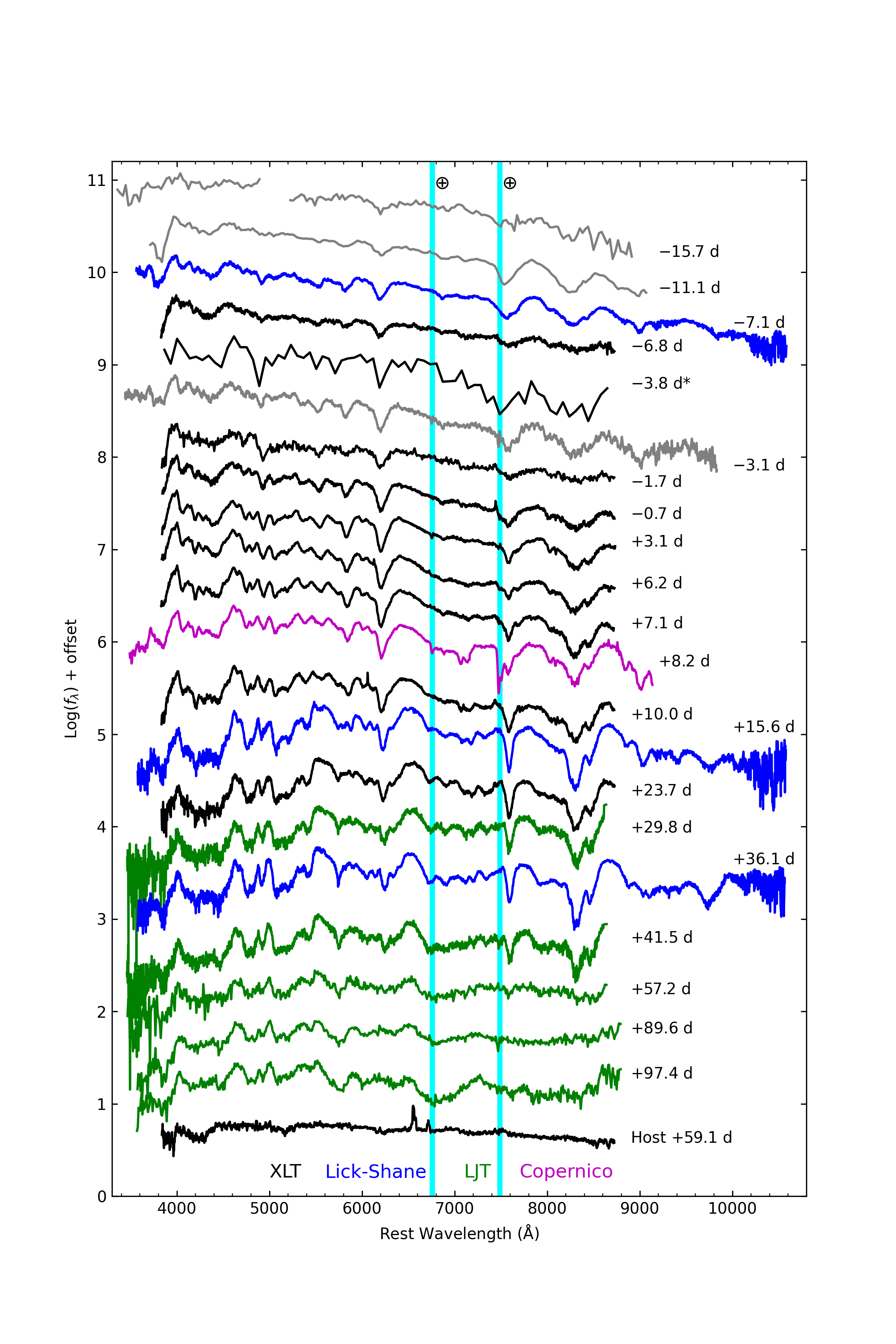}
\vspace{0.0cm}
\caption{Spectral evolution of SN~2022vqz. All spectra have been corrected for the redshift of the host galaxy ($z_{\mathrm{helio}} = 0.017$). Epochs on the right side of the spectra represent the phases in days with respect to $B$-band maximum. Colours indicate the instrument used for observations: black from the XLT, blue from the Lick Shane telescope, magenta from the Copernico telescope of Padova, and green from the LJT. Note that the $t = -3.8$~d spectrum tagged with an asterisk is of low quality and rebinned by 70~\AA. Regions of telluric correction are marked by cyan vertical bands. Telluric absorption around 7600~\AA\ was not removed from the $t = 8.2$~d spectrum. The uppermost two spectra, shown with grey solid lines, are classification spectra taken from TNS \protect\citep{2022TNSCR2778....1T,2022TNSCR2845....1M}. An LCOGT spectrum obtained at $-3.1$~d is also included, shown with a grey solid line.}
\label{figspec} \vspace{-0.0cm}
\end{figure*}

We also included two classification spectra from the Spectroscopic Classification of Astronomical Transients (SCAT; \citealt{2022PASP..134l4502T,2022TNSCR2778....1T}) and ZTF \citep{2022TNSCR2845....1M}, and one spectrum taken by Las Cumbres Observatory \citep[LCOGT; ][]{2013PASP..125.1031B} in our analysis. The spectral evolution of SN~2022vqz is shown in Figure~\ref{figspec}.

\section{Photometric Results} \label{sec:photo}

\subsection{Light Curves and Photometric Properties} \label{sec:photolc}

Figure~\ref{figlc} shows the multiband light curves of SN~2022vqz. The data reveal an early-time peak that is clearly separated from the main peak. The SN brightness increased by $\sim 0.6$ mag in the $o$ band in the first day after discovery, reaching an early peak and then dropping to a minimum in another few days, followed by a slow rise to maximum light. No secondary peak or significant shoulder can be seen in the $i$ and $R$ bands, consistent with the light curves of subluminous SNe~Ia such as the 02es-like or 91bg-like objects. 

Applying a polynomial fit to the $B$-band light curve around  maximum light indicates that SN~2022vqz reaches the peak on MJD $59863.43\pm0.48$, with $m_{B,\rm max}=16.205\pm0.053$ mag. We use the time of $B$ maximum as the reference epoch throughout this paper. The decline in the first 15 days after peak is measured as $\Delta m_{15,B}=1.33\pm0.11$ mag, slightly fast among normal SNe~Ia. 

The distance modulus of SN~2022vqz can be estimated based on the redshift of its host galaxy, which is in the Hubble flow: $z_{\rm CMB}=0.01594$ after correcting for the solar peculiar velocity. Adopting a cosmological model with $\Omega_m=0.27$, $\Omega_\Lambda=0.73$, and $H_0=73.04~{\rm km ~s^{-1}~Mpc^{-1}}$ \citep{2022ApJ...934L...7R}, the corresponding distance modulus is $m-M = 34.11\pm 0.15$ mag. With this value, the absolute peak magnitude of SN~2022vqz is $M_{B,\rm max}=-18.11\pm 0.16$ mag, after accounting for a Galactic extinction of $A_B=0.21$ mag \citep{2011ApJ...737..103S}. The extinction of the host galaxy is assumed to be negligible, considering that the SN is offset from the centre of the host and no Na~I~D absorption can be recognised in the SN spectra.

\subsection{Comparison of Light and Colour Curves} \label{sec:photocmp}

\begin{figure*}
\center
\includegraphics[angle=0,width=0.7\textwidth]{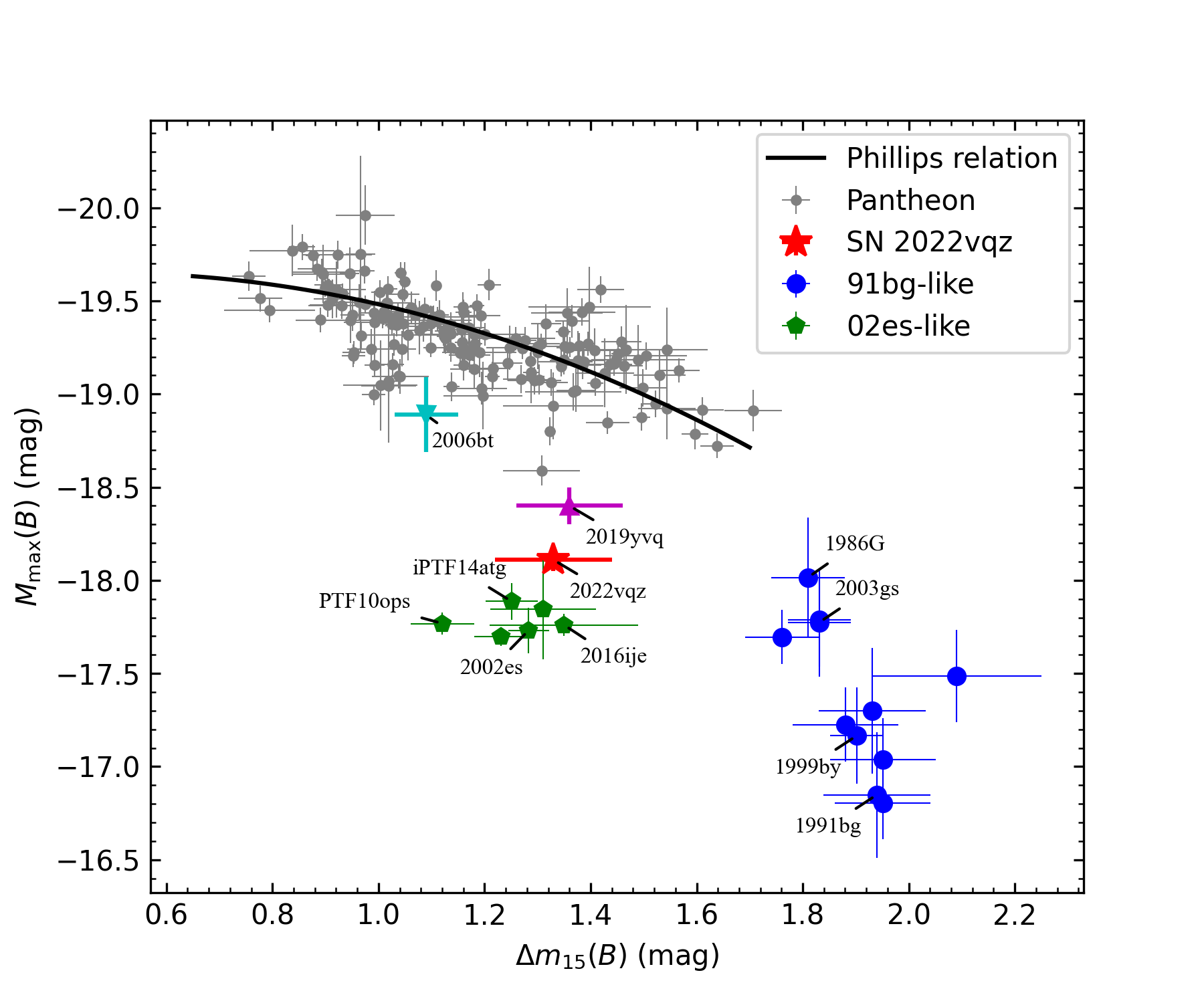}
\vspace{0.0cm}
\caption{Comparison of the $B$-band light curve decline rate and the absolute maximum magnitude for a sample of SNe~Ia. Normal SNe~Ia from the Pantheon samples \citep{2018ApJ...859..101S} are shown as grey dots. The black solid curve represents the best-fit Lira-Phillips relation \citep{1999AJ....118.1766P}. The SN~1991bg-like and SN~2002es-like samples are taken from \protect\cite{2017hsn..book..317T} and shown as blue and green dots, respectively. Several special 02es-like SNe~Ia are emphasised with different markers, including SN~2006bt \citep{2010ApJ...708.1748F}, 2019yvq \citep{2021ApJ...919..142B}, and 2022vqz (this work).}
\label{figdm15} \vspace{-0.0cm}
\end{figure*}

Figure~\ref{figdm15} shows the location of SN 2022vqz in the $\Delta m_{15,B}$--$M_{B,\rm max}$ diagram, together with those of some comparison objects including normal SNe~Ia. Its low peak luminosity and moderate decline rate fit into the category of 02es-like objects. One may note that with the discovery of some events in recent years, especially SN~2019yvq, the gap between 02es-like objects and normal SNe~Ia now seems to be filled by a continuous distribution. However, the absolute magnitude of SN~2019yvq is highly uncertain owing to large discrepancies (over 2 mag) of its host-galaxy extinction estimates from different methods \citep{2021ApJ...919..142B}. In Figure~\ref{figdm15} we adopt the value reported by \cite{2021ApJ...919..142B}.

\begin{figure*}
\center
\includegraphics[angle=0,width=1\textwidth]{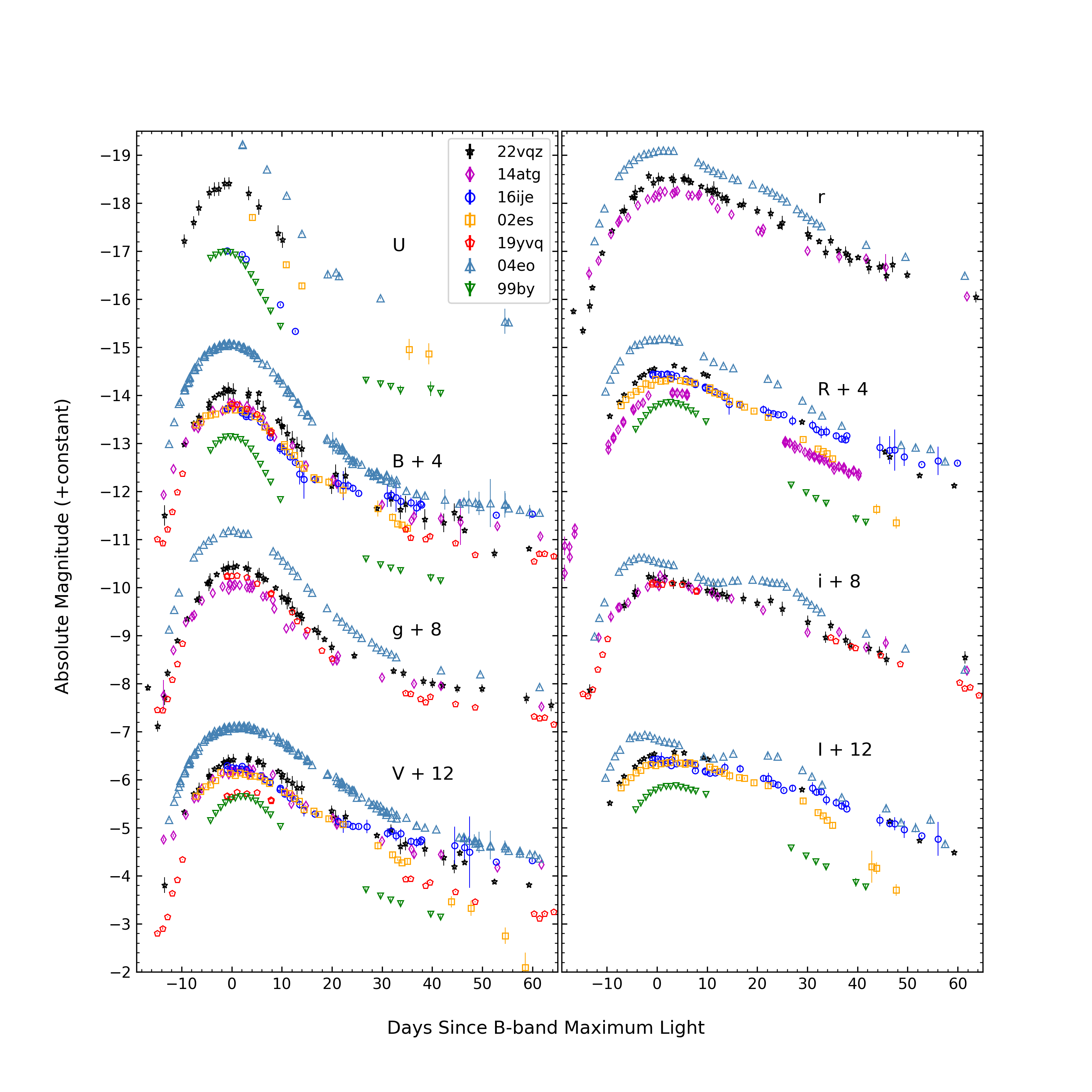}
\vspace{0.0cm}
\caption{Comparison of the multiband light curves of SN~2022vqz (black stars) with other well-observed SNe~Ia, including four 02es-like objects, one 91bg-like SN~1999by, and one normal SN~Ia 2004eo. All magnitudes are corrected for Galactic extinction.}
\label{figlccmp} \vspace{-0.0cm}
\end{figure*}

We further compare in Figure~\ref{figlccmp} the absolute $UBgV$ and $rRiI$ light curves of SN~2022vqz with those of some well-observed SNe~Ia, including 02es-like SN~2002es \citep{2012ApJ...751..142G}, SN 2016ije \citep{2023ApJ...950...17L}, SN 2019yvq \citep{2021ApJ...919..142B}, iPTF14atg \citep{2015Natur.521..328C}, 91bg-like SN~1999by \citep{2004ApJ...613.1120G}, and the normal SN~2004eo \citep{2007MNRAS.377.1531P} but with a similar $\Delta m_{15,B}$ (1.46 mag) as SN 2022vqz. 

The light-curve morphology of SN~2022vqz is overall similar to that of iPTF14atg and SN 2019yvq. We notice that SN 2022vqz and other 02es-like objects do not exhibit a rapid decline after $t \approx 1$ month when compared to SN~2002es. This indicates an additional peculiarity intrinsic to SN~2002es that is still unexplained \citep{2017hsn..book..317T}. The absolute magnitudes of SN~2022vqz are slightly brighter than all objects in the 02es-like comparison sample, while they are all bounded by subluminous 91bg-like SN~1999by and normal SN~2004eo. Note that in Figure~\ref{figlccmp} we assumed zero host extinction for all 02es-like SNe. SN~2019yvq appears not to be brighter than other 02es-like objects without applying a significant host extinction correction. Some evolution features seen in SN~2016ije, like slower $B$ and $V$ decline rates after $t\approx +20$ days and strong line blanketing in the $U$ band, are absent in SN~2022vqz. These features can be interpreted as the result of a broader line-forming region and a smaller total amount of iron-group elements (IGEs) distributed in the line of sight. A more extended line-forming region could result in stronger line blanketing at early phases, while less IGEs can reduce their absorptions in the spectra, especially at short wavelengths, thereby providing more-luminous late-time fluxes and flatter light curves in $BV$ \citep{2023ApJ...950...17L}. In contrast, SN~2022vqz should have a less extended line-forming region for IGEs, suggesting a more stratified and less mixed IGE distribution.

\begin{figure*}
\center
\includegraphics[angle=0,width=1\textwidth]{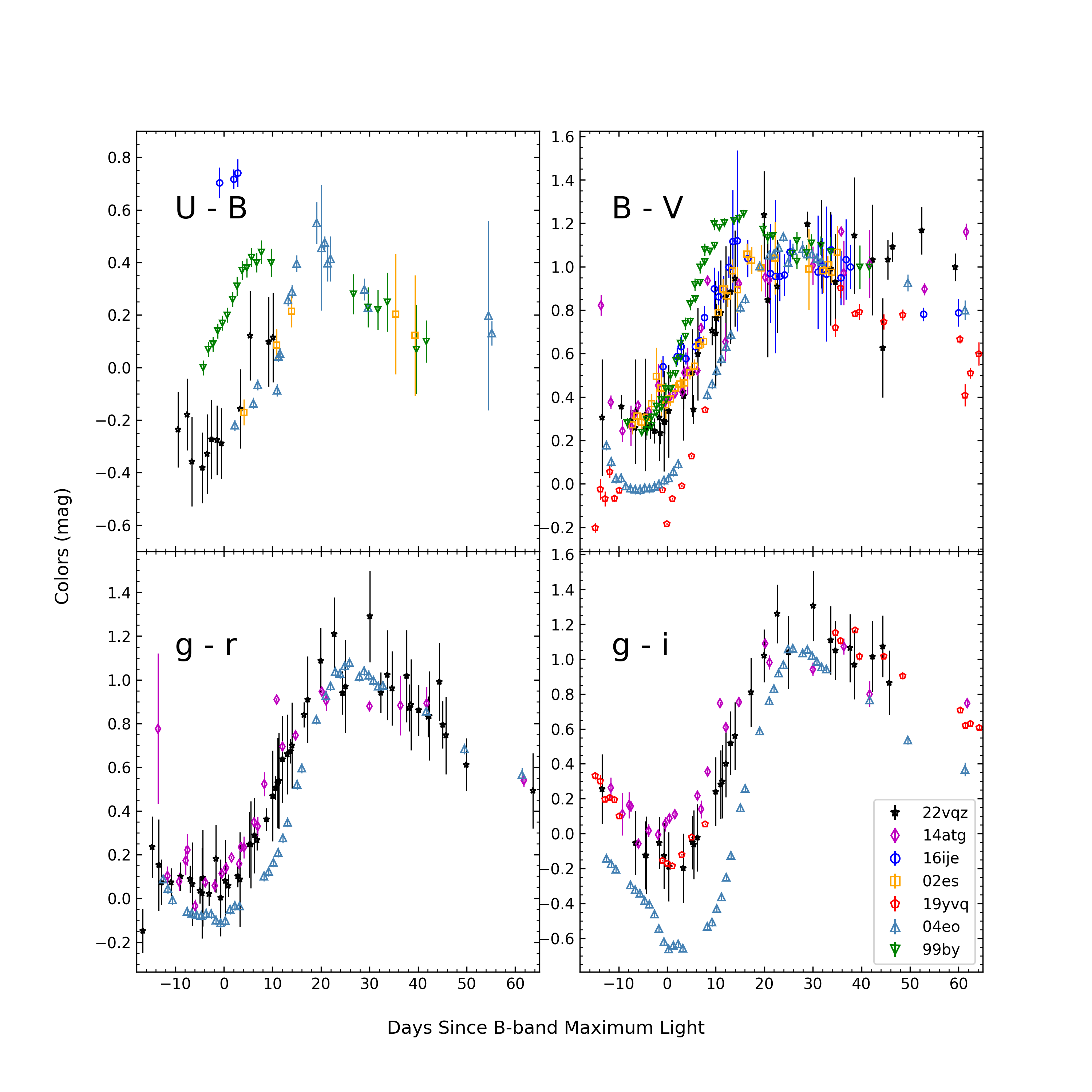}
\vspace{0.0cm}
\caption{Comparison of the $U-B$, $B-V$, $g-r$, and $g-i$ colours of SN~2022vqz with those of some well-observed SNe~Ia. All of the colour curves are corrected for their Galactic reddening. The comparison sample of SNe~Ia and the corresponding symbols are the same as in Figure~\ref{figlccmp}.}
\label{figcccmp} \vspace{-0.0cm}
\end{figure*}

Figure~\ref{figcccmp} shows the comparison of the colour evolution, with all data corrected for Galactic extinction. Overall, the colour curves of SN~2022vqz show a close resemblance to those of other 02es-like SNe, while it seems to have a more extended and rapid evolution in $g-i$  relative to iPTF14atg. In $B-V$, SN~2022vqz shows slower evolution than SN~1999by, and it is $\sim 0.2$--0.3 mag redder than SN~2004eo before $t\approx 15$ days. Immediately after the explosion, SN~2022vqz appears exceptionally blue, $g - r \approx -0.15\pm0.09$ mag at $t\approx -15$ days from maximum brightness, and it then evolves rapidly to a red peak two days later. This trend is coincident with the declining part of the early peak of SN~2022vqz, suggesting a blue colour of the early peak luminosity.

\subsection{Time of Explosion} \label{sec:photoexp}

An accurate explosion time estimation is crucial for further analysis of the explosion physics of SN 2022vqz. A nondetection was reported $\sim 1$ day before the discovery, indicating that this SN exploded not long before the first observations. The time of explosion can be estimated from the early phase of the multiband light curves by adopting a simple ``fireball'' model \citep{1999AJ....118.2675R,2011Natur.480..344N}. Assuming the early-time SN can be described by a photosphere expanding at a constant velocity, its luminosity $L$ should be proportional to $(t-t_{\rm exp})^2$, where $t_{\rm exp}$ is the time of explosion and should be same for all bands. 

\begin{figure*}
\center
\includegraphics[angle=0,width=0.5\textwidth]{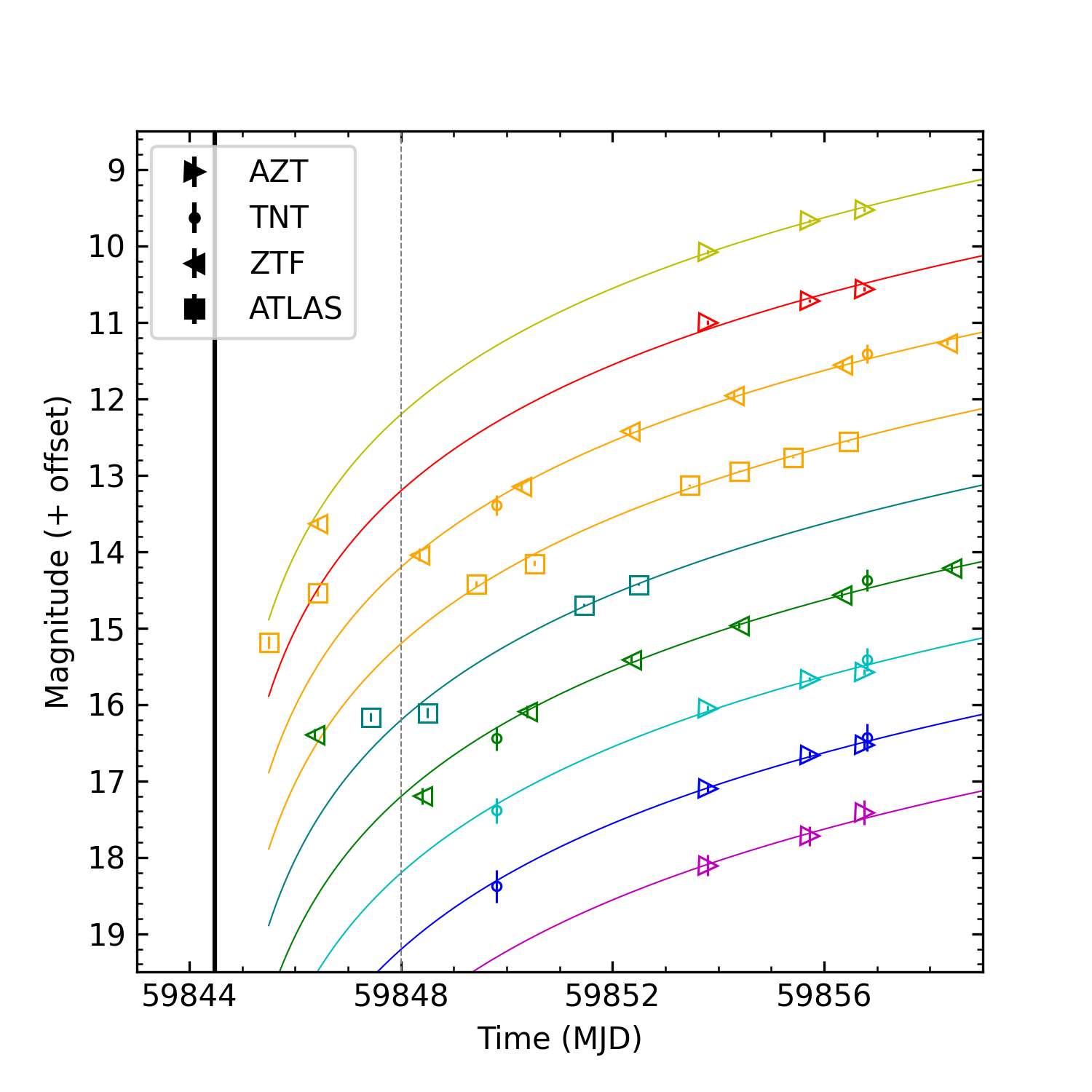}
\vspace{0.0cm}
\caption{Multiband light curves of SN~2022vqz at early phases. Colours and markers are the same as in Figure~\ref{figlc}. Solid curves are from the ``fireball'' model with explosion time MJD 59844.48. The explosion time is marked by a solid vertical line. Data before the dashed vertical line (MJD 59848) are not included in the fit to reduce the influence of the early excess. }
\label{figexp} \vspace{-0.0cm}
\end{figure*}

Note that SN~2022vqz shows an early peak which lasts for $\sim  3$ days since time of discovery and cannot be fitted by the fireball model. To avoid its influence, photometric data before MJD 59848 are excluded from the fit. The fitting method follows that described by \cite{2022MNRAS.517.4098X}, and data earlier than 5 days before $B$ maximum are used. The fitting procedure gives $t_{\rm exp}={\rm MJD}~59844.48\pm 0.20$, which coincides with the last nondetection time, MJD 59844.484. Given the uncertainty of $t_{\rm exp}$, the SN may have either not yet exploded or be too dim for detection at the latest time of nondetection. The rise time is thereby given as $t_{\rm rise}=18.63$ days. Figure~\ref{figexp} shows the fitting result.

\subsection{Bolometric Light Curve and $^{56}$Ni Mass} \label{sec:photobolo}

Based on the $UBgVrRiI$-band photometry, we adopted \texttt{SNooPy2} \citep{2011AJ....141...19B,2014ApJ...789...32B} to construct the bolometric light curve of SN~2022vqz. Since SN~2022vqz is a subluminous object, we use spectral energy distribution (SED) templates of subluminous 91bg-like SNe~Ia \citep{2002PASP..114..803N} for flux integration.

The bolometric light curve is found to reach its peak at $L\approx 5.1 \times 10^{42}~\rm{erg~s}^{-1}$ on MJD 59863.3. According to Arnett's rule \citep{1982ApJ...253..785A,2005A&A...431..423S}, the mass of synthesised radioactive $^{56}$Ni can be inferred as $M_{\rm Ni}\approx 0.25{\rm M_\odot}$. A more robust way for Ni mass estimation is to fit the full bolometric light curve using the radiation diffusion model of \cite{1982ApJ...253..785A} \citep[see also][]{2012ApJ...746..121C, 2013ApJ...773...76C}. The best fit yields the explosion time $t_0$, the initial mass of radioactive nickel $M_{\rm Ni}$, the light-curve timescale $t_{\rm lc}$, and the gamma-ray leaking timescale $t_\gamma$ as (respectively) MJD $59848.6\pm 2.2$, $0.20\pm 0.04$~M$_\odot$, $12.20\pm 3.87$~days, and $36.26\pm 4.73$~days. The explosion time inferred from the bolometric light-curve fit is about 4 days later than that from the early-time light curve, and well positioned at the minimum light between the early and main peaks, indicating a delay between the SN explosion and the emergence of the radioactive nickel-powered light curve. Such a delay was predicted as a ``dark phase'' \citep{2013ApJ...769...67P,2016ApJ...826...96P} caused by the location of the radioactive $^{56}$Ni deep within the ejecta. This delay will also result in a shorter rise time, hence a lower $^{56}$Ni mass estimation. 

The early peak bolometric luminosity was estimated as $4.5 \times 10^{41}~\rm{erg~s}^{-1}$. However, this value may be inaccurate since it only uses the ZTF $gr$-band photometry, and the rest of the flux was extrapolated from the template SED.

\section{Spectroscopic Results} \label{sec:spec}

\subsection{Spectral Evolution} \label{sec:specevol}

\begin{figure*}
\center
\includegraphics[angle=0,width=1\textwidth]{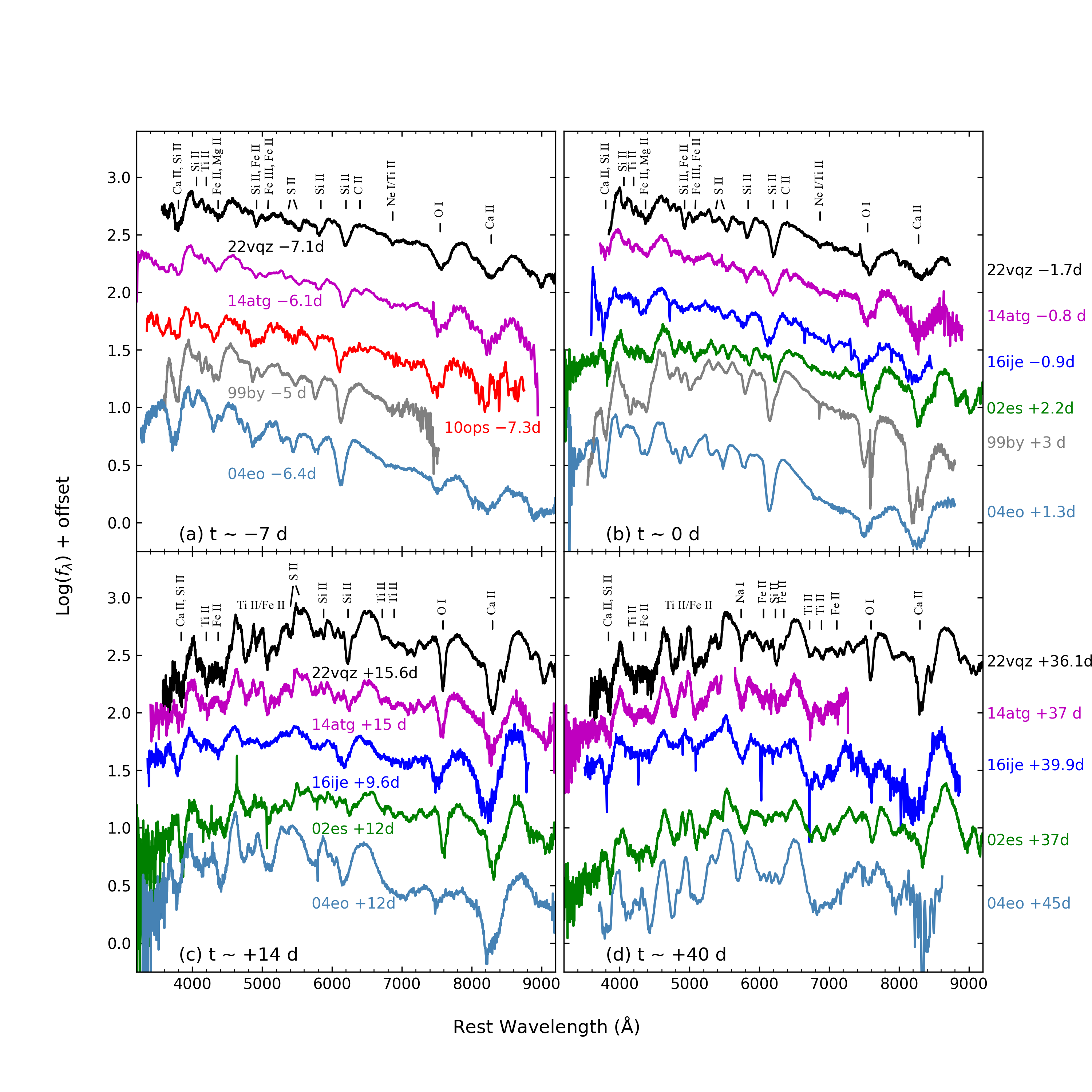}
\vspace{0.0cm}
\caption{Comparison of the spectra of SN~2022vqz (at $t \approx -7$, 0, +10, and +40~days after $B$ maximum) with those of other SNe~Ia at similar phases, including 02es-like SNe~2002es, 2016ije, PTF10ops, iPTF14atg, 91bg-like SN~1999by, and the normal SN~Ia 2004eo. All spectra have been corrected for the redshift of their host galaxy.}
\label{figspeccmp} \vspace{-0.0cm}
\end{figure*}

Figure~\ref{figspec} shows the spectral evolution of SN~2022vqz, spanning from $t\approx -15.7$ days to $+97.4$ days relative to $B$ maximum, along with a spectrum of the host galaxy. The spectral evolution generally exhibits some common properties of subluminous SNe~Ia. The Si~II $\lambda$5972 and O~I $\lambda$7774 absorptions are relatively prominent around  maximum light, implying a low ejecta temperature and a significant amount of unburnt materials. We measured the equivalent width ($W$) of the Si~II $\lambda$6355 and $\lambda$5972 features in the $-0.7$~day spectrum: $W(5972)=59 \pm 12$~\AA\ and $W(6355)=137 \pm 12$~\AA. These values make SN~2022vqz an extreme object in the ``cool'' group proposed by \cite{2006PASP..118..560B}. As explained by \cite{2006PASP..118..560B}, there is a temperature threshold of $\sim 7000$--8000~K \citep{1993A&A...268..570H,2002ApJ...568..791H}, below which the line optical depth will abruptly change owing to variations in key ionisation ratios, so the ``cool'' objects can be well separated from normal ones in the $W(6355)$--$W(5972)$ plane. The equivalent width of O~I $\lambda$7774 absorption in the same spectrum is measured to be $W(7774)=116 \pm 14$~\AA, significantly higher than those in normal SNe~Ia ($\lesssim 70$~\AA; \citealt{2016ApJ...826..211Z}). In Figure~\ref{figspeccmp}, the spectroscopic properties of SN~2022vqz are compared with those of other subluminous SNe~Ia, including 02es-like objects iPTF14atg \citep{2015Natur.521..328C}, PTF10ops \citep{2011MNRAS.418..747M}, SN~2002es \citep{2012ApJ...751..142G}, and SN 2016ije \citep{2023ApJ...950...17L}. A 91bg-like object SN~1999by \citep{2004ApJ...613.1120G} and the normal SN~Ia~2004eo \citep{2007MNRAS.377.1531P} are also included for comparison.

At $t\approx -7$ days (Figure~\ref{figspeccmp}a), the spectrum of SN~2022vqz is characterised by broad P~Cygni lines of IMEs typical among SNe~Ia: the Si~II $\lambda$6355 and $\lambda$5972 absorptions, the ``W"-shaped S~II feature around 5400~\AA, and deep absorptions of O~I $\lambda$7774 and the Ca~II near-infrared (NIR) triplet. The overall morphology is quite similar to that of iPTF14atg. However, no C~II $\lambda$6580 is detectable in SN~2022vqz, while it is strong in iPTF14atg. The weaker features of Si~II and Fe~II around 5000~\AA\ are narrower and hence less blended compared to PTF10ops or SN~2004eo.

Near maximum light (Figure~\ref{figspeccmp}b), all 02es-like objects show an absorption feature of C~II $\lambda$6580, except SN~2022vqz. The IGE lines around 5000~\AA\ are deeper and narrower than in iPTF14atg or SN~2016ije, indicating a more concentrated density distribution of the ejecta. The Ti~II feature near 4200~\AA, which is characteristic of subluminous SNe~Ia, is less developed than those of SN~2002es and 91bg-like SN~1999by at this epoch, and is more similar to iPTF14atg.

By $t\approx$ 14 days (Figure~\ref{figspeccmp}c), the spectra of SN~2022vqz resemble those of iPTF14atg and SN~2002es. The Ti~II lines have developed in the spectrum of SN~2022vqz, notably the deep trough around 4200~\AA\ and the W-shaped feature around 6800~\AA. These absorptions are less prominent or absent in SN~2016ije or SN 2004eo. The O~I and Ca~II lines remain strong, but narrower than those seen near maximum light.

At $t\approx 40$ days after maximum light (Figure~\ref{figspeccmp}d), the spectra are dominated by IGE lines. Si~II $\lambda$6355 is still pronounced in SN~2022vqz, while it is contaminated by the Fe~II lines and hardly noticeable in other objects of the comparison sample. The Si~II $\lambda$5972 absorption is also replaced by Na~I absorption, while the O~I and Ca~II lines are still persistent in the spectra.

\subsection{Ejecta Velocity} \label{sec:specvel}

\begin{figure*}
\center
\includegraphics[angle=0,width=0.7\textwidth]{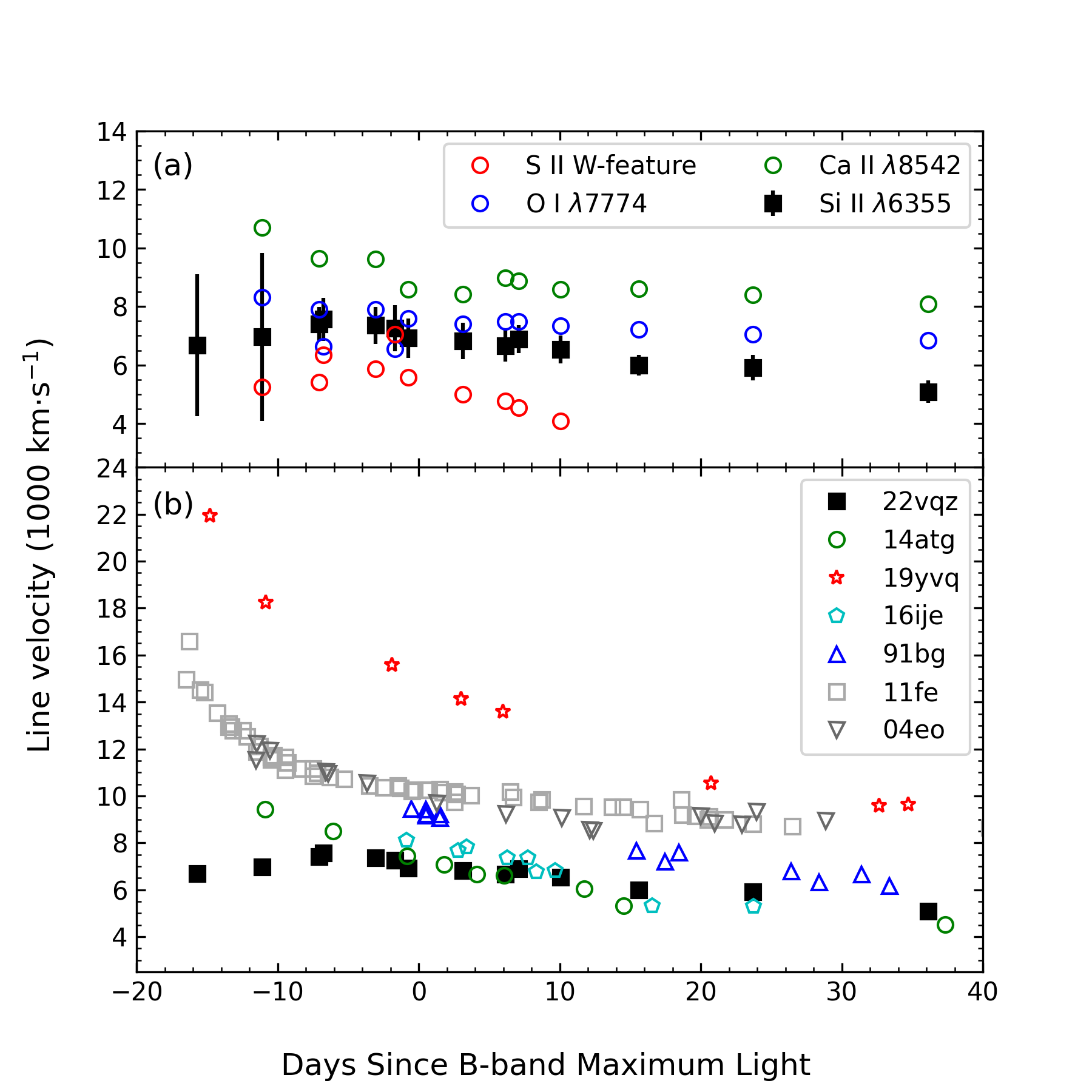}
\vspace{0.0cm}
\caption{(a)  Spectral-line velocity evolution of SN~2022vqz, measured from absorption minima of Si~II $\lambda$6355, Ca~II $\lambda$8542, O~I $\lambda$7774, and the S~II ``W'' feature ($\lambda$5468 and $\lambda\lambda$5612, 5654). (b) Evolution of the Si~II $\lambda$6355 velocity of SN~2022vqz, compared with a sample of 02es-like SNe. The velocity evolution of normal SNe~Ia is also represented by SN~2004eo and SN~2011fe. }
\label{figvel} \vspace{-0.0cm}
\end{figure*}

The expansion velocity of SN ejecta can be measured from the blueshifted absorption of P~Cygni profiles. Differences in line velocities can reveal the element distribution within the ejecta. The velocity evolution measured from absorptions of some IMEs, including S~II $\lambda\lambda$5460, 5640, O~I $\lambda$7774, Ca~II $\lambda$8542, and Si~II $\lambda$6355, appear to be exceptionally flat as shown in  Figure~\ref{figvel}(a). Despite having relatively large uncertainties, the Si~II and S~II velocities even show a global rise at early phases, conflicting with the trend that the line velocities should decrease rapidly after SN explosion as a result of the receding photosphere. This complex velocity evolution might be related to the double-peaked light curves of SN~2022vqz. Considering energy components, each results in a declining line velocity. When the first peak fades away and the second takes control, the apparent line velocity, as a weighted average of the two components, may show a temporary increase. The Si~II velocity measured around the time of maximum light is $v_{\rm Si,max}=6900\pm500~ \mathrm{km~s}^{-1}$, which is significantly lower than the typical value of $\sim 10,500$ km~s$^{-1}$ for normal SNe~Ia, and it is among the lowest values even for subluminous 91bg-like and 02es-like objects. The post-peak velocity decline of SN 2022vqz is $\dot v=-55\pm10~\mathrm{km~s^{-1}~d^{-1}}$, making it a low-velocity-gradient (LVG) object according to the classification scheme proposed by \cite{2005ApJ...623.1011B}. We do not find signatures of high-velocity features (HVFs) in SN~2022vqz, while the absence of HVFs seems to be a common property of subluminous SNe.

The velocity evolution of Si~II $\lambda$6355 is further compared to that of some other SNe~Ia, including 02es-like iPTF14atg, SN~2019yvq, SN 2016ije, subluminous SN~1991bg \citep{1996MNRAS.283....1T}, normal SN 2004eo \citep{2007MNRAS.377.1531P}, and SN 2011fe \citep{2013A&A...554A..27P,2016ApJ...820...67Z}. The latter two are overplotted to visualise the typical velocity evolution of normal SNe~Ia. The flat evolution and low velocity make SN~2022vqz stand out in Figure~\ref{figvel}(b). We also notice that, although SN~2019yvq was classified as a member of the 02es-like subclass according to its underluminous signatures \citep{2021ApJ...919..142B}, its velocity is extremely high for an 02es-like object, especially when compared with SN~2022vqz. This large diversity in ejecta velocity will be very challenging if a single mechanism is adopted to explain all 02es-like objects.

\subsection{Host-Galaxy Parameters} \label{sec:spechost}

SN~2022vqz is hosted by the face-on lenticular galaxy MCG+05-03-011 with a distance modulus of $34.11\pm0.15$~mag. The resulting $g$-band absolute magnitude is $-19.73\pm0.15$~mag and the $g-i$ colour is 1.1 mag (SDSS; \citealt{2023ApJS..267...44A}), which are typical for hosts of 02es-like SNe \citep{2015ApJ...799...52W}. Note that most 02es-like SNe~Ia tend to explode in massive, early-type galaxies with little or no star-forming activity, such as SNe 2002es \citep{2012ApJ...751..142G}, PTF10bvr, 10acdh, 10ujn \citep{2015ApJ...799...52W}, and iPTF14atg \citep{2015Natur.521..328C}. For SN 2022vqz, however, the host spectrum exhibits prominent H$\alpha$ emission, indicating relatively active star formation at least near the centre of the host galaxy. 

To determine the metallicity of the host galaxy, we measured the intensity ratio of [N~II] $\lambda$6583 and H$\alpha$ as $\log(\mathrm{[N~II]/H}\alpha)=-0.4$, and that of [O~III] $\lambda$5007 and H$\beta$ as $\log(\mathrm{[O~III]/H}\beta)=-0.1$. These values can be converted to a metallicity estimate of $12+\log(\mathrm{O/H})=8.64$ using an empirical relationship (\citealt{2008ApJ...681.1183K}, Eq. A9). We also used \texttt{Firefly} \citep{2017MNRAS.472.4297W} to fit the host spectrum with combinations of single-burst stellar population models, and obtained a stellar mass of $\log({M/\mathrm{M}_\odot})=9.79$, an age of 4.05~Gyr, and a metallicity $[\mathrm{Z/H}]=0.14$~dex. The metallicities inferred from the two methods are roughly consistent and comparable to the solar metallicity. Thus, we conclude that the host of SN~2022vqz is a medium-mass, solar-metallicity, star-forming galaxy.

\section{Discussion} \label{sec:dis}

\subsection{Double-Detonation Model} \label{sec:dispeak}

Early-time photometry provides a distinct perspective and is crucial for discriminating different explosion mechanisms of SNe~Ia \citep{2018ApJ...861...78M}. Recent statistical studies based on ZTF samples found that $\sim 20$--30\% of normal SNe~Ia show some early-time flux excesses \citep{2022MNRAS.512.1317D,2022MNRAS.513.3035M}. There is also large diversity in brightness, colours, timescales, and light-curve morphologies of these excesses. Different physical models have been proposed to explain these features, such as interaction with a nondegenerate companion star \citep{2010ApJ...708.1025K}, outward mixing of radioactive $^{56}$Ni \citep{2013ApJ...769...67P,2016ApJ...826...96P}, detonation of a surface He layer \citep{2019ApJ...873...84P}, and interaction with CSM \citep{2016ApJ...826...96P,2021ApJ...909..209P}. The rate of early excess in normal SNe~Ia is related to the actual population of different physical channels. Even if all the normal SNe~Ia are from the SD scenario, an early excess from companion interaction could only be detected in $\sim 10$\% of the events owing to the viewing-angle effect \citep{2010ApJ...708.1025K,2022arXiv220707681B}.

Obtaining high-quality early-time photometry necessitates early discovery, rapid classification, and high-cadence follow-up photometry of young SNe~Ia, which require both extensive efforts and good luck. Early excess emission has been observed in detail only for a handful of normal SNe~Ia, such as SN~2017cbv \citep{2017ApJ...845L..11H}, SN~2018oh \citep{2019ApJ...870L...1D,2019ApJ...870...12L,2019ApJ...870...13S}, SN~2019np \citep{2022MNRAS.514.3541S}, and SN~2023bee \citep{2023arXiv230503779W,2023ApJ...953L..15H}. In the future, with specially designed robotic facilities dedicated to high-cadence, multiband observations of young, nearby SNe~Ia, we expect more such events to be observed and analysed in detail.

Although early-time flux excesses are relatively rare among normal SNe~Ia, a series of recent studies reveals that the 02es-like SNe~Ia tend to show them: see iPTF14atg \citep{2015Natur.521..328C}, iPTF14dpk \citep{2016ApJ...832...86C,2018ApJ...865..149J}, SN~2016jhr \citep{2017Natur.550...80J}, SN~2019yvq \citep{2020ApJ...898...56M,2021ApJ...919..142B}, SN~2022ywc \citep{2023arXiv230806019S}, and SN~2022vqz (this work). This generic feature supports the idea that the 02es-like subclass likely originates from a single channel. However, the diversity in other aspects, including ejecta velocity, peak luminosity, and existence of early line blanketing, challenges the potential progenitor scenario and explosion mechanism. One possible scenario is the double-detonation model, in which the ignition of a surface He layer detonates the underlying C-O WD \citep{2019ApJ...873...84P}. The burning of the He layer will leave radioactive IGEs in the outermost ejecta and produce an early-time excess. This model was adopted to explain the early excesses of SNe~2016jhr and 2019yvq.

\begin{figure*}
\center
\includegraphics[angle=0,width=0.9\textwidth]{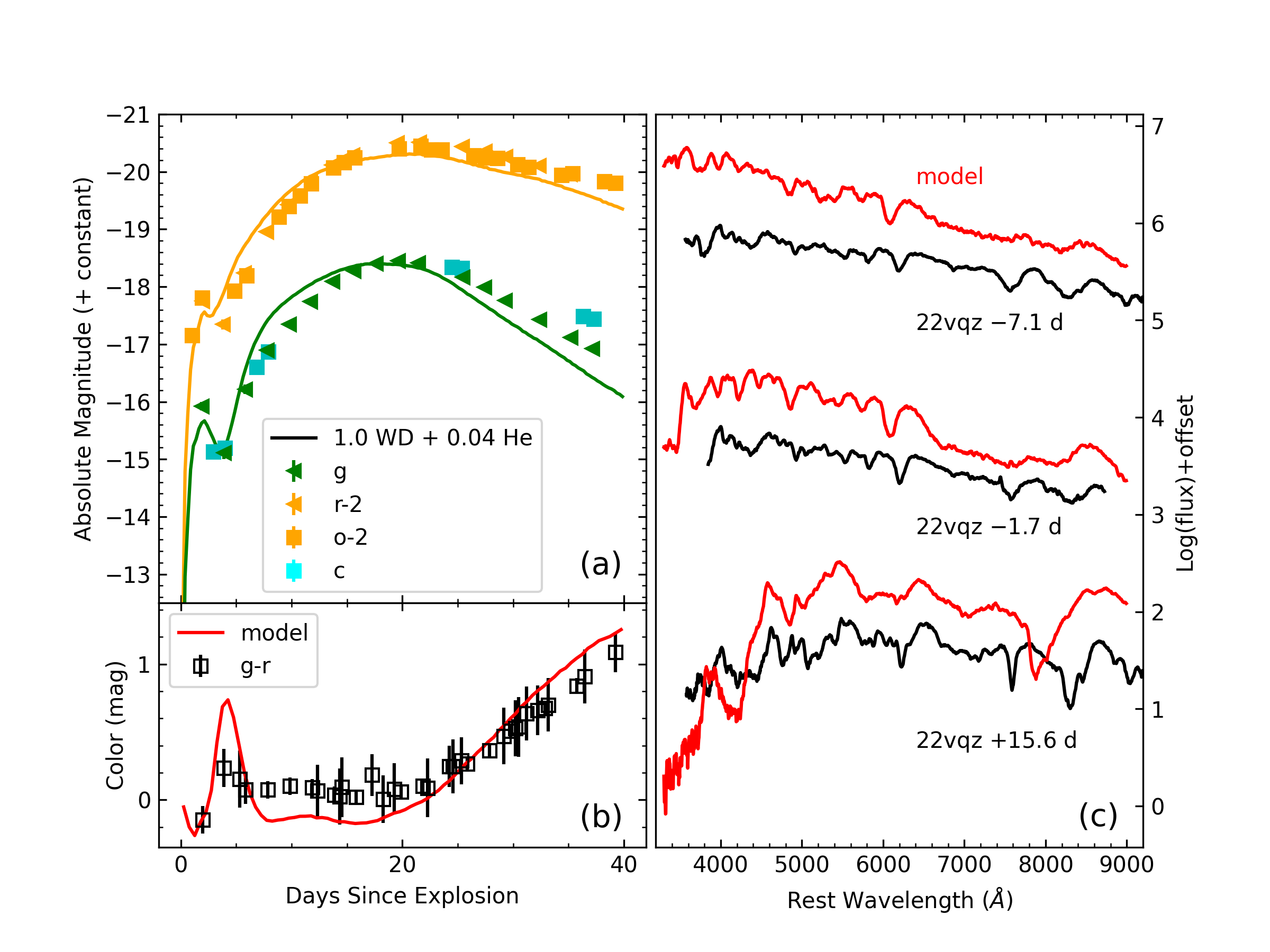}
\vspace{0.0cm}
\caption{Photometry and spectra of SN~2022vqz compared to the double-detonation model of a 1.0~M$_\odot$ WD with a 0.04~M$_\odot$ He shell. (a) $g$-band and $r$-band absolute light curves. (b) $g-r$ colour. (c) Spectra at different epochs. The model light curves are shifted by 0.8 mag to match the peaks of the observed light curves.}
\label{figpolin} \vspace{-0.0cm}
\end{figure*}

We compared the observations of SN~2022vqz with a grid of double-detonation models \citep{2019ApJ...873...84P}. The grid contains models with WD masses ranging from 0.6 to 1.3~${\rm M}_\odot$ and He-shell masses varying between 0.01 and 0.1~${\rm M}_\odot$. The explosion epoch is taken as the value inferred in Section~\ref{sec:photoexp}. The best-fit model has a WD mass of 1.0~${\rm M}_\odot$ and a He shell of 0.04~${\rm M}_\odot$. Comparisons between this model and the observed $gr$-band light curves, the $g-r$ colour, and the spectra are shown in Figure~\ref{figpolin}. 

The early $gr$ peaks of SN~2022vqz appear to be well reproduced, and the relative strengths of the early peak to the main peak is also satisfactory. In particular, the ``red bump'' in $g-r$ colour $\sim 4$ days after the explosion is overpredicted by the model. The general trend of the subsequent light-curve evolution is also reproduced, with minor differences. The model light curves are faster, bluer at early epochs and redder at later times, and overluminous by a factor of $\sim 2$ (0.8 mag). All of this makes the model light curves more consistent with those of fast decliners among normal SNe~Ia rather than 02es-like events.

Most spectral features can be reproduced by the double-detonation model, but the predicted velocities are higher than the observed ones by 3000--5000~km~s$^{-1}$. This is not unexpected given that SN~2022vqz is a peculiar object with an extremely low velocity, as described in Section~\ref{sec:specvel}. No carbon can be detected in the model and the observed spectra. O~I $\lambda$7774 is underpredicted, especially in the $+14.6$ day spectrum. Moreover, the model predicts a strong high-velocity Ca~II feature, but this is not observed in the $+14.6$ day spectrum.

Although the double-detonation model is capable of reproducing the early-time peak, all of the other discrepancies, especially those in spectral velocities, are difficult to reconcile. Reducing the WD mass to 0.9~${\rm M}_\odot$ yields lower luminosity and ejecta velocities, providing a better match with the observations. On the other hand, changing the above parameters would give a worse match with the overall light-curve and colour-curve evolution, and the velocities of Si~II and Ca~II features would still be unreasonably high.

\subsection{O-Ne WD} \label{sec:disonemg}

It is commonly accepted that SNe~Ia are explosions of C-O WDs. Assuming the WD is composed of equal parts of C and O, and is fully burnt into iron-peak elements, the released specific energy is $1.55\times 10^{51}$~erg~${\rm M}_\odot^{-1}$ \citep{1992ApJ...392...35B}. This energy will be turned into kinetic energy of the ejecta, after compensating the binding energy of the WD. Since heavier WDs are more compact and have a larger specific binding energy, the thermonuclear explosion of a sub-$M_{\rm Ch}$ WD is expected to result in an SN with higher ejecta velocity than that of a $M_{\rm Ch}$ WD, assuming the same progenitor abundance and burning efficiency. This trend is beneficial in explaining high-velocity SNe like SN~2019ein \citep{2022MNRAS.517.4098X} and SN 2019yvq \citep{2021ApJ...919..142B}, but causes problems for the low-velocity SN~2022vqz.

Reducing the burning efficiency will result in lower energy yields and slower ejecta velocities, leaving signatures of unburnt elements in the spectra. Most of the subluminous 02es-like SNe show a noticeable C~II $\lambda$6580 feature around  maximum light, indicating low-efficiency burning of C-O WDs. However, the spectra of SN~2022vqz do not reveal any trace of carbon. Thus, we tentatively consider a possibility that the progenitor WD is made of oxygen and neon (O-Ne). By replacing $^{12}$C with $^{20}$Ne, the specific fusion energy reduces to  $1.21\times 10^{51}$~erg~${\rm M}_\odot^{-1}$ and hence causes a decrease in the ejecta velocity by $\sim 2000$~km~s$^{-1}$, which could potentially explain the low velocity observed in SN~2022vqz.

\begin{figure*}
\center
\includegraphics[angle=0,width=0.7\textwidth]{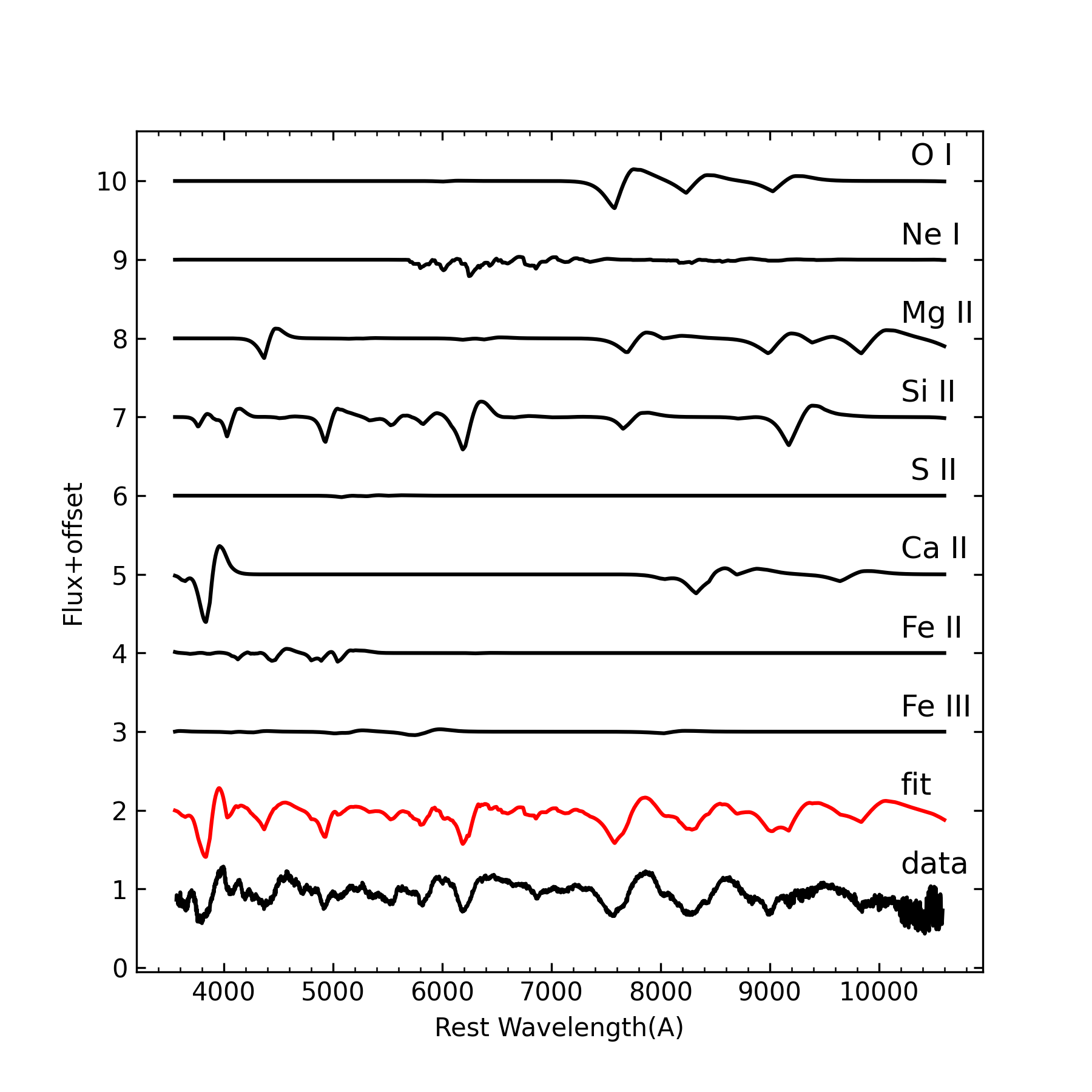}
\vspace{0.0cm}
\caption{\texttt{SYNAPPS} fit to the $-7.1$ day Lick spectrum (at bottom). Spectral features from individual ions are plotted above the synthesised spectrum (red).  }
\label{figone} \vspace{-0.0cm}
\end{figure*}

We further use \texttt{SYNAPPS} \citep{2011PASP..123..237T} to identify the potential signatures of unburnt Ne in the spectra of SN 2022vqz.  Figure~\ref{figone} shows the decomposition of the spectrum synthesised by \texttt{SYNAPPS} compared to the $t=-7.1$ day spectrum of SN~2022vqz. We noticed that the shallow absorption features around 6800~\AA\ and 7100~\AA\ could be well fit by Ne~I. These features are relatively common in subluminous SNe (see Figure~\ref{figspeccmp}). Since burning of a C-O WD may also produce Ne, this cannot be treated as an exclusive signature of an O-Ne WD progenitor. The absorption  around 6800~\AA\ might be also attributed to Ti~II. However, the 6800~\AA\ Ti~II feature usually emerges at later epochs and is accompanied by another Ti~II feature at $\sim 6600$~\AA\ which is invisible in this spectrum. Thus, we still identify the 6800~\AA\ absorption feature in the $t=-7.1$ day spectrum as Ne~I rather than Ti~II.

The electron-degenerate cores made of O-Ne are believed to result from carbon burning in ``heavy-weight'' intermediate-mass stars (i.e., $8{\rm M}_\odot \lesssim M\lesssim 11{\rm M}_\odot$), and they are expected to collapse into neutron stars through electron capture \citep{1980PASJ...32..303M}. However, O-Ne WDs can be ignited at lower central densities if a tiny amount of carbon ($\sim 1.5$\%) exists in their cores \citep{2005A&A...435..231G}, leading to thermonuclear SNe~Ia. This tiny amount of carbon can be explained as the remnant of an off-centre carbon-burning process which does not propagate all the way to the centre. \cite{2016A&A...589A..38B} and \cite{2016ApJ...832...13W} examined the delayed detonation of such hybrid C-O-Ne cores. They found that the produced ejecta are characterised by slightly lower masses of $^{56}$Ni and substantially less kinetic energy than those of normal C-O WDs, making this kind of event an interesting candidate for the subluminous class of SNe~Ia.

 While the C-O-Ne model can simultaneously explain the deficient of C and extremely low velocity of SN~2022vqz, it does not naturally reproduce the early excess emission, unless combined with the double-detonation model (see Section~\ref{sec:dispeak}). Most of the observables can be well reproduced or explained by detonation of a C-O-Ne core triggered by a thick He shell. The companion star could either be a helium star or another degenerate He WD. \cite{2015A&A...580A.118M} studied a set of double-detonation models of $\sim 1.2$~M$_\odot$ O-Ne WDs. The ejecta structure and spectral morphology of O-Ne models are similar to those of the C-O models of similar WD masses, but the resultant explosions are less energetic with peak luminosity being fainter by $\sim 0.2$ mag in B and ejecta velocity slower by $\sim 3000$~km~s$^{-1}$, respectively.

However, it should be noted that the minimal mass of a C-O-Ne core is $\sim 1.08$~M$_\odot$, which is the required mass for a C-O core to start off-centre carbon burning to Ne \citep{2000MNRAS.315..543H,2015A&A...580A.118M}. This value is slightly larger than the 1.0~M$_\odot$ inferred from model fitting in Section~\ref{sec:dispeak} and will predict a brighter peak magnitude. It is not clear whether this mass excess can be compensated by the subluminous nature of O-Ne core explosions compared with C-O cores. Viewing angle may also affect the observed luminosity since double detonation is intrinsically asymmetric. Further analysis requires a more thorough numerical examination of double detonation of C-O-Ne WDs, exploring the full mass range and viewing angle effects of such events.

\subsection{CSM Interaction} \label{sec:discsm}

The early excesses of some 02es-like events are attributed to interaction between ejecta and CSM. \cite{2016MNRAS.459.4428K} suggest that interaction with a nonspherical CSM may be able to account for the early UV emission in iPTF14atg. \cite{2023arXiv230806019S} also attribute the strong early-time peak observed in SN~2022ywc to CSM interaction, and rule out other scenarios like surface helium detonation, surface Ni distribution, or companion interaction. 

\begin{figure*}
\center
\includegraphics[angle=0,width=0.7\textwidth]{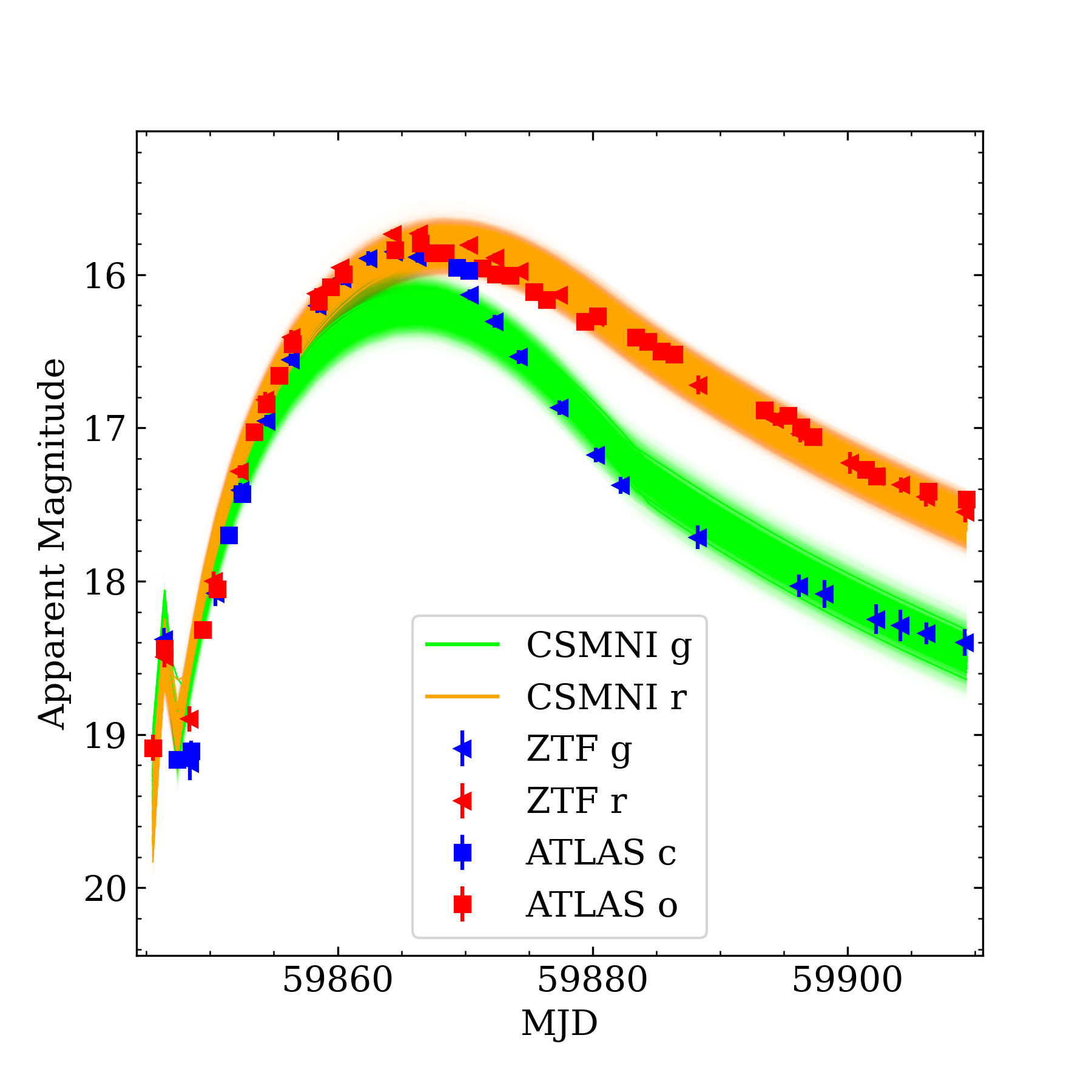}
\vspace{0.0cm}
\caption{\texttt{MOSFiT} fitting of the ZTF $gr$-band and ATLAS $co$-band photometry of SN~2022vqz. Shaded regions are clumps of light curves from the \texttt{CSMNI} model.}
\label{figmosfit} \vspace{-0.0cm}
\end{figure*}

Following the methodology of \cite{2023arXiv230806019S}, we use the \texttt{CSMNI} model of Modular Open Source Fitter for Transients (\texttt{MOSFiT}; \citealt{2018ApJS..236....6G,2017ApJ...850...55N}) to fit the $gr$-band photometry of SN~2022vqz. The best-fit parameters are $E_{\rm k}=0.51^{+0.15}_{-0.07}\times 10^{51}$~erg, $M_{\rm CSM}=0.0056^{+0.0042}_{-0.0015}~{\rm M}_\odot$, $M_{\rm ej}=0.87^{+0.11}_{-0.06}~{\rm M}_\odot$, $R_0=2.6^{+2.1}_{-1.6}\times 10^{14}$~cm, and $t_{\rm exp}={\rm MJD}~59844.75^{+0.08}_{-0.11}$, where $E_{\rm k}$ is the ejecta kinetic energy, $M_{\rm CSM}$ is the CSM mass, $M_{\rm ej}$ is the ejecta mass, $R_0$ is the inner radius of the CSM shell, and $t_{\rm exp}$ is the explosion time. Figure~\ref{figmosfit} shows the $gr$-band light curves of SN~2022vqz compared with those from the \texttt{CSMNI} model.

The \texttt{CSMNI} model reproduces the timescale and shape of the early peak very well, and it also provides a good fit to the main bulk of the light curves, with a slightly fainter predicted $g$-band peak. The estimated explosion time is $\sim 0.3$ days later than that inferred from the fireball model in Section~\ref{sec:photoexp}, but within the quoted uncertainty. The ejecta mass is significantly less than $M_{\rm Ch}$, making near-$M_{\rm Ch}$ models like the delayed detonation unfavourable. This model requires a CSM distribution at $\sim 3\times 10^{14}$~cm, similar to that of SN~2022ywc, but with a mass of about one order of magnitude lower (i.e., $\sim 0.006~{\rm M}_\odot$).

The CSM could originate from interactions between two merging WDs, specifically the violent merger scenario. This scenario was favoured for multiple 02es-like events, including SNe~2002es, 2010lp, PTF10ops, and iPTF14atg. The secondary WD can be disrupted by the primary, forming a centrifugally supported disc \citep{2007MNRAS.380..933Y,2012ApJ...748...35S}, or eject disc-originated material (DOM; e.g., disk winds or jets) along the polar directions \citep{2015MNRAS.447.2803L}. Simulations of WD mergers indicate that the mass of CSM could reach $\sim 10^{-4}$--$10^{-2}~{\rm M}_\odot$ for discs formed by tidal tails \citep{2013ApJ...772....1R,2014MNRAS.438...14D}, or $\sim 10^{-2}$--$10^{-1}~{\rm M}_\odot$ for disc wind-driven DOM \citep{2019ApJ...872L...7L}. The nonspherical configuration of CSM may lead to a few 02es-like events that have very bright early excesses, as SN~2022ywc, if the viewing angle aligns with the polar axis \citep{2023arXiv230806019S}. The large parameter space and asymmetry of the violent merger scenario may account for the observed diversity in luminosity and ejecta velocity of 02es-like objects.

\subsection{ Other Models for the Early Excess} \label{sec:disother}

Some other mechanisms could also produce early excess emission in SNe~Ia. One is the collision between SN ejecta and a nondegenerate companion star \citep{2010ApJ...708.1025K,2019ApJ...870L...1D}; a bright UV/optical excess will be produced when the SN ejecta reach the surface of the companion. The separation between the SN and its companion should be comparable to the Roche-lobe radius of the companion star, $a/R_{\rm companion}=2$--3, where $a$ is the separation. For typical velocities of SN and radius of main-sequence stars, the interaction flux should appear a few minutes after the SN explosion, or a couple of hours in the case of a red-giant companion \citep{2010ApJ...708.1025K}.

Assuming the density profile of the outer SN ejecta follows a power law, $\rho \propto r^{-10}$, the isotropic equivalent collision-powered luminosity ($L_\mathrm{c}$) and effective temperature ($T_\mathrm{eff}$) can be analytically estimated as:

\begin{equation}
L_\mathrm{c} \propto a M_\mathrm{ej}^{1/4}v_\mathrm{ej}^{7/4}\kappa_e^{-3/4}t^{-1/2}\, ,
\end{equation}
\begin{equation}
T_\mathrm{eff} \propto a^{1/4}\kappa_e^{-35/36}t^{-37/72}\, ,
\end{equation}

where $a$ is the distance between the WD and its companion, $M_\mathrm{ej}$ is the ejecta mass, $v_\mathrm{ej}$ is the ejecta velocity, $\kappa_e$ is the effective opacity in optical, and $t$ is the time after explosion, respectively. Assuming the early-time SED can be treated as a black-body radiation, multiband model photometry and colour evolution can be constructed from $L_\mathrm{c}$ and $T_\mathrm{eff}$.

\begin{figure*}
\center
\includegraphics[angle=0,width=1.0\textwidth]{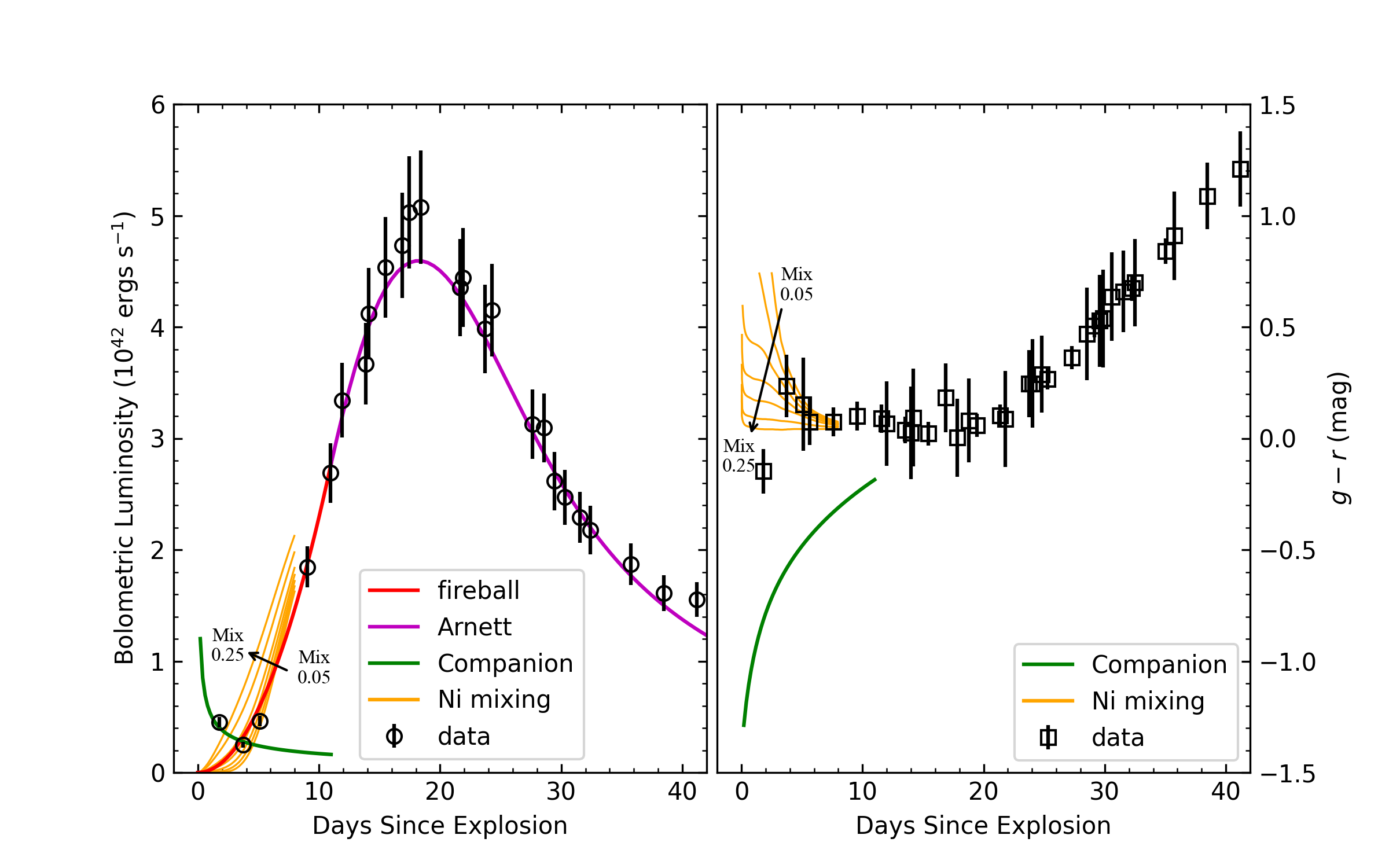}
\vspace{0.0cm}
\caption{Bolometric light curve and $g-r$ colour curve of SN~2022vqz (black symbols), compared with companion interaction model (green lines) and Ni mixing models (orange lines). Ni mixing model curves along two black arrows are generated by boxcar parameters of 0.05, 0.075, 0.1, 0.125, 0.15, 0.2, and 0.25, respectively; higher values indicate higher levels of mixing. The bolometric light curve is also fitted by a combination of the fireball model (red) and the radiation diffusion model (magenta) of \protect\cite{1982ApJ...253..785A}.}
\label{figother} \vspace{-0.0cm}
\end{figure*}

We fit the observed bolometric light curve and $g-r$ colour evolution to the analytic model. As the early-time bolometric light curve is not sufficiently sampled, to reduce the free parameters in the fitting, the opacity $\kappa_e$ is set to $0.2$~cm$^2$~g$^{-1}$ according to the assumption made by \cite{2010ApJ...708.1025K}, whereas $M_\mathrm{ej}$ and $v_\mathrm{ej}$ are fixed to $M_\mathrm{ch}$ and 7000~km~s$^{-1}$, respectively. Figure~\ref{figother} shows the best-fit companion-collision model with separation $a=4.3$\,R$_\odot$. While the decreasing part of the early bolometric peak is well fitted, the model curve does not show an initial increase before the early peak as observed in ATLAS $o$ and ZTF $r$ bands. Moreover, the model $g-r$ colour curve is too blue, and lacks a red bump as seen in the observed data of SN 2022vqz.

Another possible explanation for the early flux excesses is  $^{56}$Ni mixing to the outermost layers of the SN ejecta \citep{2013ApJ...769...67P,2022MNRAS.514.3541S}. The emission from the outer $^{56}$Ni can escape earlier than that from the deeper one, and will cause an early excess in first few days after the SN explosion. \cite{2016ApJ...826...96P} examined a series of theoretical models with different levels of Ni mixing, measured by the width of a boxcar smoothing routine on the Ni distribution curve (0.05 $\le$ boxcar $\le$ 0.25). Their model bolometric and colour curves are also overplotted in Figure~\ref{figother} for comparison. When the level of mixing increases, the model bolometric light curve shows a more rapid increase at early times, and the colour curve becomes flatter and bluer. However, neither the early bolometric peak nor the red-bump feature of SN~2022vqz can be reproduced by the Ni-mixing models. The absence of a separate early peak is not surprising, since the outer Ni in these models is from mixing of a concentrated distribution, and the mass fraction of Ni still decreases monotonically from centre to the surface. In contrast, the outermost Ni in the double-detonation model is produced by surface He, thus $^{56}$Ni can be more congregated in the outermost region of ejecta and produce a clear early peak.

The main strengths and weaknesses of the considered models in this paper are summarised in Table~\ref{tabmodels}.

\section{Conclusion} \label{sec:sum}

In this paper, we presented extensive photometric and spectroscopic observations of SN~2022vqz, a peculiar SN~Ia located in a solar-metallicity, medium-mass, star-forming galaxy. SN~2022vqz has an absolute $B$-band peak magnitude of $M_{B,\rm max}=-18.11\pm 0.16$ mag and a post-peak decline rate of $\Delta m_{15,B}=1.33\pm 0.11$ mag, which fits well into the subclass of subluminous 02es-like SNe~Ia. The bolometric light curve of SN~2022vqz is constructed from multiband photometry and fitted to the radiation diffusion model of \cite{1982ApJ...253..785A}, implying a $^{56}$Ni mass of $M_{\rm Ni}=0.20\pm 0.04$~M$_\odot$.

The light curves and spectra of SN~2022vqz resemble those of the classical (i.e., subluminous, low ejecta velocity) 02es-like object iPTF14atg very well. However, while lower ejecta velocity indicates a less energetic explosion and lower burning efficiency, no C is detected in the spectra of SN~2022vqz, in contrast to the prominent C~II absorption seen in pre-peak and maximum-light spectra of iPTF14atg.

SN~2022vqz also displays a clear blue early-time peak within $\sim 4$ days after the explosion, confirming the recently established trend that most, if not all, 02es-like events have early blue/UV excesses. We discussed the possible origin of this early excess. A sub-$M_{\rm Ch}$ double-detonation model of $\sim 0.9$--$1.0~{\rm M}_\odot$ WD + $0.04~{\rm M}_\odot$ He shell can well reproduce the strength and epoch of the early peak, but it also predicts a ``red bump'' after the early peak that is significantly stronger than the observed data. Double-detonation models generally leave little or no C in spectra, which is consistent with SN~2022vqz. However, the model spectral features are significantly blueshifted compared to the observed ones, overpredicting the ejecta velocity by 3000--5000~km~s$^{-1}$. This discrepancy may be reconciled by considering a hybrid C-O-Ne WD model, which releases lower specific fusion energy when ignited and produces SN~Ia explosions with slightly lower $^{56}$Ni masses and substantially slower ejecta velocities.

We also use the code \texttt{MOSFiT} to explore the CSM interaction scenario as an alternative explanation for the early peak. A CSM mass of $\sim 0.006~{\rm M}_\odot$ at $\sim 3\times 10^{14}$~cm is required to reproduce the early peak. This amount of CSM could arise from interactions of WDs in a violent merger scenario, which has been proposed for several 02es-like events. The large amount of asymmetry and extent of parameter space may account for the observed diversity in luminosity, ejecta velocity, and strength of the early excess in the 02es-like subclass. However, since the CSM is formed by disruption of a secondary C-O WD, it should be C-O-rich and produce C signatures in spectra of the SN. The lack of a C signature in SN~2022vqz is still a challenge for this model. 

Additional models like SN ejecta interaction with a nondegenerate companion and outward Ni mixing are also examined. None can simultaneously reproduce the morphology of the bolometric light curve and colour evolution of the early peak. Moreover, they do not provide additional clues for the cool, subluminous, and slow-expansion nature of the 02es-like SNe~Ia; thus, we disfavour these models for SN~2022vqz.

In conclusion, we suggest that double detonation of a low-mass ($\lesssim 1.1$~M$_\odot$) C-O-Ne WD is the most viable model for SN~2022vqz. However, there are still some tensions in peak luminosity and synthesised $^{56}$Ni mass that require further modeling to resolve. SN~2022vqz extends the diversity of the 02es-like subclass with its extremely low ejecta velocity and absence of carbon features. Additional observations of this SN at nebular phases would be important for discriminating between different possible progenitor scenarios. More efforts in exploring the observational properties of progenitors that previously drew less attention, like WDs made of C-O-Ne, will likely help in determining the possible origin of this rare subclass of SNe~Ia. Three-dimensional numerical simulations that explore the whole parameter space and viewing angles of double detonation and violent merger models, although challenging, will also help in understanding the diversity in 02es-like events. For now, the sample size of 02es-like SNe~Ia is still limited. Future surveys that concentrate on quick response for young transients and early-time multiwavelength photometric observations may greatly enrich the sample, providing better constraints on the origin of the seemingly universal early-time excesses of 02es-like SNe~Ia.

\section*{Acknowledgements}

This work is supported by the National Natural Science Foundation of China (NSFC; grants 12288102, 12033003, and 11633002), the Scholar Program of Beijing Academy of Science and Technology (DZ:BS202002), and the Tencent Xplorer Prize. Y.-Z. Cai and J.-J. Zhang are supported by the International Centre of Supernovae, Yunnan Key Laboratory (No. 202302AN360001). Y.-Z. Cai is supported by NSFC grant 12303054. A.V.F.'s group at U.C. Berkeley received financial support from the Christopher R. Redlich Fund, Briggs and Kathleen Wood (T.G.B. is a Wood Specialist in Astronomy), Alan Eustace (W.Z. is a Eustace Specialist in Astronomy), and many individual donors. A.P., A.R., and L.T. acknowledge support from the PRIN-INAF 2022, ``Shedding light on the nature of gap transients: from the observations to the models.''

A major upgrade of the Kast spectrograph on the Shane 3\,m telescope at Lick Observatory was made possible through generous gifts from William and Marina Kast as well as the Heising-Simons Foundation. Research at Lick Observatory is partially supported by a generous gift from Google.
Observations at AZT-22 of the Maidanak Observatory were supported by Uzbekistan's Ministry of Innovative Development (grant A-FA-2021-36). This work is partially based on observations collected with the Copernico 1.82\,m telescope and the Schmidt 67/92 telescope (Asiago Mount Ekar, Italy) INAF - Osservatorio Astronomico di Padova. The ZTF forced-photometry service was funded under Heising-Simons Foundation grant 12540303 (PI: M. Graham).
This work makes use of the NASA/IPAC Extragalactic Database (NED), which is funded by NASA and operated by the California Institute of Technology.

  This work has made use of data from the Asteroid Terrestrial-impact Last Alert System (ATLAS) project. The Asteroid Terrestrial-impact Last Alert System (ATLAS) project is primarily funded to search for near-Earth objects through NASA grants NN12AR55G, 80NSSC18K0284, and 80NSSC18K1575; byproducts of the NEO search include images and catalogs from the survey area. This work was partially funded by Kepler/K2 grant J1944/80NSSC19K0112 and HST GO-15889, and by STFC grants ST/T000198/1 and ST/S006109/1. The ATLAS science products have been made possible through the contributions of the University of Hawaii Institute for Astronomy, the Queen's University Belfast, the Space Telescope Science Institute, the South African Astronomical Observatory, and The Millennium Institute of Astrophysics (MAS), Chile.

\section*{Data Availability}

The photometric data underlying this article are available in the article, and the spectroscopic data will be available in the Weizmann Interactive Supernova Data Repository (WISeREP) at https://www.wiserep.org/object/21648 .



\bibliographystyle{mnras}
\bibliography{references} 

\begin{thebibliography}{}
\makeatletter
\relax
\def\mn@urlcharsother{\let\do\@makeother \do\$\do\&\do\#\do\^\do\_\do\%\do\~}
\def\mn@doi{\begingroup\mn@urlcharsother \@ifnextchar [ {\mn@doi@} {\mn@doi@[]}}
\def\mn@doi@[#1]#2{\def\@tempa{#1}\ifx\@tempa\@empty \href {http://dx.doi.org/#2} {doi:#2}\else \href {http://dx.doi.org/#2} {#1}\fi \endgroup}
\def\mn@eprint#1#2{\mn@eprint@#1:#2::\@nil}
\def\mn@eprint@arXiv#1{\href {http://arxiv.org/abs/#1} {{\tt arXiv:#1}}}
\def\mn@eprint@dblp#1{\href {http://dblp.uni-trier.de/rec/bibtex/#1.xml} {dblp:#1}}
\def\mn@eprint@#1:#2:#3:#4\@nil{\def\@tempa {#1}\def\@tempb {#2}\def\@tempc {#3}\ifx \@tempc \@empty \let \@tempc \@tempb \let \@tempb \@tempa \fi \ifx \@tempb \@empty \def\@tempb {arXiv}\fi \@ifundefined {mn@eprint@\@tempb}{\@tempb:\@tempc}{\expandafter \expandafter \csname mn@eprint@\@tempb\endcsname \expandafter{\@tempc}}}

\bibitem[\protect\citeauthoryear{{Almeida} et~al.,}{{Almeida} et~al.}{2023}]{2023ApJS..267...44A}
{Almeida} A.,  et~al., 2023, \mn@doi [\apjs] {10.3847/1538-4365/acda98}, \href {https://ui.adsabs.harvard.edu/abs/2023ApJS..267...44A} {267, 44}

\bibitem[\protect\citeauthoryear{{Arnett}}{{Arnett}}{1982}]{1982ApJ...253..785A}
{Arnett} W.~D.,  1982, \mn@doi [\apj] {10.1086/159681}, \href {https://ui.adsabs.harvard.edu/abs/1982ApJ...253..785A} {253, 785}

\bibitem[\protect\citeauthoryear{{Benetti} et~al.,}{{Benetti} et~al.}{2005}]{2005ApJ...623.1011B}
{Benetti} S.,  et~al., 2005, \mn@doi [\apj] {10.1086/428608}, \href {https://ui.adsabs.harvard.edu/abs/2005ApJ...623.1011B} {623, 1011}

\bibitem[\protect\citeauthoryear{{Bloom} et~al.,}{{Bloom} et~al.}{2012}]{2012ApJ...744L..17B}
{Bloom} J.~S.,  et~al., 2012, \mn@doi [\apjl] {10.1088/2041-8205/744/2/L17}, \href {https://ui.adsabs.harvard.edu/abs/2012ApJ...744L..17B} {744, L17}

\bibitem[\protect\citeauthoryear{{Branch}}{{Branch}}{1992}]{1992ApJ...392...35B}
{Branch} D.,  1992, \mn@doi [\apj] {10.1086/171401}, \href {https://ui.adsabs.harvard.edu/abs/1992ApJ...392...35B} {392, 35}

\bibitem[\protect\citeauthoryear{{Branch}, {Fisher}  \& {Nugent}}{{Branch} et~al.}{1993}]{1993AJ....106.2383B}
{Branch} D.,  {Fisher} A.,   {Nugent} P.,  1993, \mn@doi [\aj] {10.1086/116810}, \href {https://ui.adsabs.harvard.edu/abs/1993AJ....106.2383B} {106, 2383}

\bibitem[\protect\citeauthoryear{{Branch} et~al.,}{{Branch} et~al.}{2006}]{2006PASP..118..560B}
{Branch} D.,  et~al., 2006, \mn@doi [\pasp] {10.1086/502778}, \href {https://ui.adsabs.harvard.edu/abs/2006PASP..118..560B} {118, 560}

\bibitem[\protect\citeauthoryear{{Bravo}, {Gil-Pons}, {Guti{\'e}rrez}  \& {Doherty}}{{Bravo} et~al.}{2016}]{2016A&A...589A..38B}
{Bravo} E.,  {Gil-Pons} P.,  {Guti{\'e}rrez} J.~L.,   {Doherty} C.~L.,  2016, \mn@doi [\aap] {10.1051/0004-6361/201527861}, \href {https://ui.adsabs.harvard.edu/abs/2016A&A...589A..38B} {589, A38}

\bibitem[\protect\citeauthoryear{{Brown} et~al.,}{{Brown} et~al.}{2013}]{2013PASP..125.1031B}
{Brown} T.~M.,  et~al., 2013, \mn@doi [\pasp] {10.1086/673168}, \href {https://ui.adsabs.harvard.edu/abs/2013PASP..125.1031B} {125, 1031}

\bibitem[\protect\citeauthoryear{{Burke} et~al.,}{{Burke} et~al.}{2021}]{2021ApJ...919..142B}
{Burke} J.,  et~al., 2021, \mn@doi [\apj] {10.3847/1538-4357/ac126b}, \href {https://ui.adsabs.harvard.edu/abs/2021ApJ...919..142B} {919, 142}

\bibitem[\protect\citeauthoryear{{Burke} et~al.,}{{Burke} et~al.}{2022}]{2022arXiv220707681B}
{Burke} J.,  et~al., 2022, \mn@doi [arXiv e-prints] {10.48550/arXiv.2207.07681}, \href {https://ui.adsabs.harvard.edu/abs/2022arXiv220707681B} {p. arXiv:2207.07681}

\bibitem[\protect\citeauthoryear{{Burns} et~al.,}{{Burns} et~al.}{2011}]{2011AJ....141...19B}
{Burns} C.~R.,  et~al., 2011, \mn@doi [\aj] {10.1088/0004-6256/141/1/19}, \href {https://ui.adsabs.harvard.edu/abs/2011AJ....141...19B} {141, 19}

\bibitem[\protect\citeauthoryear{{Burns} et~al.,}{{Burns} et~al.}{2014}]{2014ApJ...789...32B}
{Burns} C.~R.,  et~al., 2014, \mn@doi [\apj] {10.1088/0004-637X/789/1/32}, \href {https://ui.adsabs.harvard.edu/abs/2014ApJ...789...32B} {789, 32}

\bibitem[\protect\citeauthoryear{{Cao} et~al.,}{{Cao} et~al.}{2015}]{2015Natur.521..328C}
{Cao} Y.,  et~al., 2015, \mn@doi [\nat] {10.1038/nature14440}, \href {https://ui.adsabs.harvard.edu/abs/2015Natur.521..328C} {521, 328}

\bibitem[\protect\citeauthoryear{{Cao}, {Kulkarni}, {Gal-Yam}, {Papadogiannakis}, {Nugent}, {Masci}  \& {Bue}}{{Cao} et~al.}{2016}]{2016ApJ...832...86C}
{Cao} Y.,  {Kulkarni} S.~R.,  {Gal-Yam} A.,  {Papadogiannakis} S.,  {Nugent} P.~E.,  {Masci} F.~J.,   {Bue} B.~D.,  2016, \mn@doi [\apj] {10.3847/0004-637X/832/1/86}, \href {https://ui.adsabs.harvard.edu/abs/2016ApJ...832...86C} {832, 86}

\bibitem[\protect\citeauthoryear{{Chatzopoulos}, {Wheeler}  \& {Vinko}}{{Chatzopoulos} et~al.}{2012}]{2012ApJ...746..121C}
{Chatzopoulos} E.,  {Wheeler} J.~C.,   {Vinko} J.,  2012, \mn@doi [\apj] {10.1088/0004-637X/746/2/121}, \href {https://ui.adsabs.harvard.edu/abs/2012ApJ...746..121C} {746, 121}

\bibitem[\protect\citeauthoryear{{Chatzopoulos}, {Wheeler}, {Vinko}, {Horvath}  \& {Nagy}}{{Chatzopoulos} et~al.}{2013}]{2013ApJ...773...76C}
{Chatzopoulos} E.,  {Wheeler} J.~C.,  {Vinko} J.,  {Horvath} Z.~L.,   {Nagy} A.,  2013, \mn@doi [\apj] {10.1088/0004-637X/773/1/76}, \href {https://ui.adsabs.harvard.edu/abs/2013ApJ...773...76C} {773, 76}

\bibitem[\protect\citeauthoryear{{Dan}, {Rosswog}, {Br{\"u}ggen}  \& {Podsiadlowski}}{{Dan} et~al.}{2014}]{2014MNRAS.438...14D}
{Dan} M.,  {Rosswog} S.,  {Br{\"u}ggen} M.,   {Podsiadlowski} P.,  2014, \mn@doi [\mnras] {10.1093/mnras/stt1766}, \href {https://ui.adsabs.harvard.edu/abs/2014MNRAS.438...14D} {438, 14}

\bibitem[\protect\citeauthoryear{{Deckers} et~al.,}{{Deckers} et~al.}{2022}]{2022MNRAS.512.1317D}
{Deckers} M.,  et~al., 2022, \mn@doi [\mnras] {10.1093/mnras/stac558}, \href {https://ui.adsabs.harvard.edu/abs/2022MNRAS.512.1317D} {512, 1317}

\bibitem[\protect\citeauthoryear{{Dimitriadis} et~al.,}{{Dimitriadis} et~al.}{2019}]{2019ApJ...870L...1D}
{Dimitriadis} G.,  et~al., 2019, \mn@doi [\apjl] {10.3847/2041-8213/aaedb0}, \href {https://ui.adsabs.harvard.edu/abs/2019ApJ...870L...1D} {870, L1}

\bibitem[\protect\citeauthoryear{{Ehgamberdiev}}{{Ehgamberdiev}}{2018}]{2018NatAs...2..349E}
{Ehgamberdiev} S.,  2018, \mn@doi [Nature Astronomy] {10.1038/s41550-018-0459-3}, \href {https://ui.adsabs.harvard.edu/abs/2018NatAs...2..349E} {2, 349}

\bibitem[\protect\citeauthoryear{{Fan}, {Bai}, {Zhang}, {Wang}, {Chang}, {Xin}  \& {Zhang}}{{Fan} et~al.}{2015}]{2015RAA....15..918F}
{Fan} Y.-F.,  {Bai} J.-M.,  {Zhang} J.-J.,  {Wang} C.-J.,  {Chang} L.,  {Xin} Y.-X.,   {Zhang} R.-L.,  2015, \mn@doi [Research in Astronomy and Astrophysics] {10.1088/1674-4527/15/6/014}, \href {https://ui.adsabs.harvard.edu/abs/2015RAA....15..918F} {15, 918}

\bibitem[\protect\citeauthoryear{{Filippenko}}{{Filippenko}}{1982}]{1982PASP...94..715F}
{Filippenko} A.~V.,  1982, \mn@doi [\pasp] {10.1086/131052}, \href {https://ui.adsabs.harvard.edu/abs/1982PASP...94..715F} {94, 715}

\bibitem[\protect\citeauthoryear{{Filippenko}}{{Filippenko}}{1997}]{1997ARA&A..35..309F}
{Filippenko} A.~V.,  1997, \mn@doi [\araa] {10.1146/annurev.astro.35.1.309}, \href {https://ui.adsabs.harvard.edu/abs/1997ARA&A..35..309F} {35, 309}

\bibitem[\protect\citeauthoryear{{Filippenko} et~al.,}{{Filippenko} et~al.}{1992}]{1992AJ....104.1543F}
{Filippenko} A.~V.,  et~al., 1992, \mn@doi [\aj] {10.1086/116339}, \href {https://ui.adsabs.harvard.edu/abs/1992AJ....104.1543F} {104, 1543}

\bibitem[\protect\citeauthoryear{{Foley}, {Narayan}, {Challis}, {Filippenko}, {Kirshner}, {Silverman}  \& {Steele}}{{Foley} et~al.}{2010}]{2010ApJ...708.1748F}
{Foley} R.~J.,  {Narayan} G.,  {Challis} P.~J.,  {Filippenko} A.~V.,  {Kirshner} R.~P.,  {Silverman} J.~M.,   {Steele} T.~N.,  2010, \mn@doi [\apj] {10.1088/0004-637X/708/2/1748}, \href {https://ui.adsabs.harvard.edu/abs/2010ApJ...708.1748F} {708, 1748}

\bibitem[\protect\citeauthoryear{{Fremling}}{{Fremling}}{2022}]{2022TNSTR2766....1F}
{Fremling} C.,  2022, Transient Name Server Discovery Report, \href {https://ui.adsabs.harvard.edu/abs/2022TNSTR2766....1F} {2022-2766, 1}

\bibitem[\protect\citeauthoryear{{Ganeshalingam} et~al.,}{{Ganeshalingam} et~al.}{2012}]{2012ApJ...751..142G}
{Ganeshalingam} M.,  et~al., 2012, \mn@doi [\apj] {10.1088/0004-637X/751/2/142}, \href {https://ui.adsabs.harvard.edu/abs/2012ApJ...751..142G} {751, 142}

\bibitem[\protect\citeauthoryear{{Garnavich} et~al.,}{{Garnavich} et~al.}{2004}]{2004ApJ...613.1120G}
{Garnavich} P.~M.,  et~al., 2004, \mn@doi [\apj] {10.1086/422986}, \href {https://ui.adsabs.harvard.edu/abs/2004ApJ...613.1120G} {613, 1120}

\bibitem[\protect\citeauthoryear{{Guillochon}, {Nicholl}, {Villar}, {Mockler}, {Narayan}, {Mandel}, {Berger}  \& {Williams}}{{Guillochon} et~al.}{2018}]{2018ApJS..236....6G}
{Guillochon} J.,  {Nicholl} M.,  {Villar} V.~A.,  {Mockler} B.,  {Narayan} G.,  {Mandel} K.~S.,  {Berger} E.,   {Williams} P. K.~G.,  2018, \mn@doi [\apjs] {10.3847/1538-4365/aab761}, \href {https://ui.adsabs.harvard.edu/abs/2018ApJS..236....6G} {236, 6}

\bibitem[\protect\citeauthoryear{{Gunn} et~al.,}{{Gunn} et~al.}{2006}]{2006AJ....131.2332G}
{Gunn} J.~E.,  et~al., 2006, \mn@doi [\aj] {10.1086/500975}, \href {https://ui.adsabs.harvard.edu/abs/2006AJ....131.2332G} {131, 2332}

\bibitem[\protect\citeauthoryear{{Guti{\'e}rrez}, {Canal}  \& {Garc{\'\i}a-Berro}}{{Guti{\'e}rrez} et~al.}{2005}]{2005A&A...435..231G}
{Guti{\'e}rrez} J.,  {Canal} R.,   {Garc{\'\i}a-Berro} E.,  2005, \mn@doi [\aap] {10.1051/0004-6361:20042254}, \href {https://ui.adsabs.harvard.edu/abs/2005A&A...435..231G} {435, 231}

\bibitem[\protect\citeauthoryear{{Guy} et~al.,}{{Guy} et~al.}{2007}]{2007A&A...466...11G}
{Guy} J.,  et~al., 2007, \mn@doi [\aap] {10.1051/0004-6361:20066930}, \href {https://ui.adsabs.harvard.edu/abs/2007A&A...466...11G} {466, 11}

\bibitem[\protect\citeauthoryear{{Hachinger}, {Mazzali}, {Taubenberger}, {Pakmor}  \& {Hillebrandt}}{{Hachinger} et~al.}{2009}]{2009MNRAS.399.1238H}
{Hachinger} S.,  {Mazzali} P.~A.,  {Taubenberger} S.,  {Pakmor} R.,   {Hillebrandt} W.,  2009, \mn@doi [\mnras] {10.1111/j.1365-2966.2009.15403.x}, \href {https://ui.adsabs.harvard.edu/abs/2009MNRAS.399.1238H} {399, 1238}

\bibitem[\protect\citeauthoryear{{Hoeflich}, {Mueller}  \& {Khokhlov}}{{Hoeflich} et~al.}{1993}]{1993A&A...268..570H}
{Hoeflich} P.,  {Mueller} E.,   {Khokhlov} A.,  1993, \aap, \href {https://ui.adsabs.harvard.edu/abs/1993A&A...268..570H} {268, 570}

\bibitem[\protect\citeauthoryear{{H{\"o}flich}, {Gerardy}, {Fesen}  \& {Sakai}}{{H{\"o}flich} et~al.}{2002}]{2002ApJ...568..791H}
{H{\"o}flich} P.,  {Gerardy} C.~L.,  {Fesen} R.~A.,   {Sakai} S.,  2002, \mn@doi [\apj] {10.1086/339063}, \href {https://ui.adsabs.harvard.edu/abs/2002ApJ...568..791H} {568, 791}

\bibitem[\protect\citeauthoryear{{Hoogendam}, {Shappee}, {Brown}, {Tucker}, {Ashall}  \& {Piro}}{{Hoogendam} et~al.}{2023}]{2023arXiv230911563H}
{Hoogendam} W.~B.,  {Shappee} B.~J.,  {Brown} P.~J.,  {Tucker} M.~A.,  {Ashall} C.,   {Piro} A.~L.,  2023, \mn@doi [arXiv e-prints] {10.48550/arXiv.2309.11563}, \href {https://ui.adsabs.harvard.edu/abs/2023arXiv230911563H} {p. arXiv:2309.11563}

\bibitem[\protect\citeauthoryear{{Hosseinzadeh} et~al.,}{{Hosseinzadeh} et~al.}{2017}]{2017ApJ...845L..11H}
{Hosseinzadeh} G.,  et~al., 2017, \mn@doi [\apjl] {10.3847/2041-8213/aa8402}, \href {https://ui.adsabs.harvard.edu/abs/2017ApJ...845L..11H} {845, L11}

\bibitem[\protect\citeauthoryear{{Hosseinzadeh} et~al.,}{{Hosseinzadeh} et~al.}{2023}]{2023ApJ...953L..15H}
{Hosseinzadeh} G.,  et~al., 2023, \mn@doi [\apjl] {10.3847/2041-8213/ace7c0}, \href {https://ui.adsabs.harvard.edu/abs/2023ApJ...953L..15H} {953, L15}

\bibitem[\protect\citeauthoryear{{Howell} et~al.,}{{Howell} et~al.}{2006}]{2006Natur.443..308H}
{Howell} D.~A.,  et~al., 2006, \mn@doi [\nat] {10.1038/nature05103}, \href {https://ui.adsabs.harvard.edu/abs/2006Natur.443..308H} {443, 308}

\bibitem[\protect\citeauthoryear{{Hoyle} \& {Fowler}}{{Hoyle} \& {Fowler}}{1960}]{1960ApJ...132..565H}
{Hoyle} F.,  {Fowler} W.~A.,  1960, \mn@doi [\apj] {10.1086/146963}, \href {https://ui.adsabs.harvard.edu/abs/1960ApJ...132..565H} {132, 565}

\bibitem[\protect\citeauthoryear{{Huang}, {Li}, {Wang}, {Shang}, {Zhang}, {Hu}, {Qiu}  \& {Jiang}}{{Huang} et~al.}{2012}]{2012RAA....12.1585H}
{Huang} F.,  {Li} J.-Z.,  {Wang} X.-F.,  {Shang} R.-C.,  {Zhang} T.-M.,  {Hu} J.-Y.,  {Qiu} Y.-L.,   {Jiang} X.-J.,  2012, \mn@doi [Research in Astronomy and Astrophysics] {10.1088/1674-4527/12/11/012}, \href {https://ui.adsabs.harvard.edu/abs/2012RAA....12.1585H} {12, 1585}

\bibitem[\protect\citeauthoryear{{Hurley}, {Pols}  \& {Tout}}{{Hurley} et~al.}{2000}]{2000MNRAS.315..543H}
{Hurley} J.~R.,  {Pols} O.~R.,   {Tout} C.~A.,  2000, \mn@doi [\mnras] {10.1046/j.1365-8711.2000.03426.x}, \href {https://ui.adsabs.harvard.edu/abs/2000MNRAS.315..543H} {315, 543}

\bibitem[\protect\citeauthoryear{{Iben} \& {Tutukov}}{{Iben} \& {Tutukov}}{1984}]{1984ApJS...54..335I}
{Iben} I. J.,  {Tutukov} A.~V.,  1984, \mn@doi [\apjs] {10.1086/190932}, \href {https://ui.adsabs.harvard.edu/abs/1984ApJS...54..335I} {54, 335}

\bibitem[\protect\citeauthoryear{{Jha}, {Riess}  \& {Kirshner}}{{Jha} et~al.}{2007}]{2007ApJ...659..122J}
{Jha} S.,  {Riess} A.~G.,   {Kirshner} R.~P.,  2007, \mn@doi [\apj] {10.1086/512054}, \href {https://ui.adsabs.harvard.edu/abs/2007ApJ...659..122J} {659, 122}

\bibitem[\protect\citeauthoryear{{Jiang} et~al.,}{{Jiang} et~al.}{2017}]{2017Natur.550...80J}
{Jiang} J.-A.,  et~al., 2017, \mn@doi [\nat] {10.1038/nature23908}, \href {https://ui.adsabs.harvard.edu/abs/2017Natur.550...80J} {550, 80}

\bibitem[\protect\citeauthoryear{{Jiang}, {Doi}, {Maeda}  \& {Shigeyama}}{{Jiang} et~al.}{2018}]{2018ApJ...865..149J}
{Jiang} J.-a.,  {Doi} M.,  {Maeda} K.,   {Shigeyama} T.,  2018, \mn@doi [\apj] {10.3847/1538-4357/aadb9a}, \href {https://ui.adsabs.harvard.edu/abs/2018ApJ...865..149J} {865, 149}

\bibitem[\protect\citeauthoryear{{Kasen}}{{Kasen}}{2010}]{2010ApJ...708.1025K}
{Kasen} D.,  2010, \mn@doi [\apj] {10.1088/0004-637X/708/2/1025}, \href {https://ui.adsabs.harvard.edu/abs/2010ApJ...708.1025K} {708, 1025}

\bibitem[\protect\citeauthoryear{{Kewley} \& {Ellison}}{{Kewley} \& {Ellison}}{2008}]{2008ApJ...681.1183K}
{Kewley} L.~J.,  {Ellison} S.~L.,  2008, \mn@doi [\apj] {10.1086/587500}, \href {https://ui.adsabs.harvard.edu/abs/2008ApJ...681.1183K} {681, 1183}

\bibitem[\protect\citeauthoryear{{Kromer} et~al.,}{{Kromer} et~al.}{2013}]{2013ApJ...778L..18K}
{Kromer} M.,  et~al., 2013, \mn@doi [\apjl] {10.1088/2041-8205/778/1/L18}, \href {https://ui.adsabs.harvard.edu/abs/2013ApJ...778L..18K} {778, L18}

\bibitem[\protect\citeauthoryear{{Kromer} et~al.,}{{Kromer} et~al.}{2016}]{2016MNRAS.459.4428K}
{Kromer} M.,  et~al., 2016, \mn@doi [\mnras] {10.1093/mnras/stw962}, \href {https://ui.adsabs.harvard.edu/abs/2016MNRAS.459.4428K} {459, 4428}

\bibitem[\protect\citeauthoryear{{Leibundgut} et~al.,}{{Leibundgut} et~al.}{1993}]{1993AJ....105..301L}
{Leibundgut} B.,  et~al., 1993, \mn@doi [\aj] {10.1086/116427}, \href {https://ui.adsabs.harvard.edu/abs/1993AJ....105..301L} {105, 301}

\bibitem[\protect\citeauthoryear{{Levanon} \& {Soker}}{{Levanon} \& {Soker}}{2019}]{2019ApJ...872L...7L}
{Levanon} N.,  {Soker} N.,  2019, \mn@doi [\apjl] {10.3847/2041-8213/ab0285}, \href {https://ui.adsabs.harvard.edu/abs/2019ApJ...872L...7L} {872, L7}

\bibitem[\protect\citeauthoryear{{Levanon}, {Soker}  \& {Garc{\'\i}a-Berro}}{{Levanon} et~al.}{2015}]{2015MNRAS.447.2803L}
{Levanon} N.,  {Soker} N.,   {Garc{\'\i}a-Berro} E.,  2015, \mn@doi [\mnras] {10.1093/mnras/stu2580}, \href {https://ui.adsabs.harvard.edu/abs/2015MNRAS.447.2803L} {447, 2803}

\bibitem[\protect\citeauthoryear{{Li} et~al.,}{{Li} et~al.}{2019}]{2019ApJ...870...12L}
{Li} W.,  et~al., 2019, \mn@doi [\apj] {10.3847/1538-4357/aaec74}, \href {https://ui.adsabs.harvard.edu/abs/2019ApJ...870...12L} {870, 12}

\bibitem[\protect\citeauthoryear{{Li} et~al.,}{{Li} et~al.}{2023}]{2023ApJ...950...17L}
{Li} Z.,  et~al., 2023, \mn@doi [\apj] {10.3847/1538-4357/accde3}, \href {https://ui.adsabs.harvard.edu/abs/2023ApJ...950...17L} {950, 17}

\bibitem[\protect\citeauthoryear{{Liu}, {R{\"o}pke}  \& {Han}}{{Liu} et~al.}{2023}]{2023RAA....23h2001L}
{Liu} Z.-W.,  {R{\"o}pke} F.~K.,   {Han} Z.,  2023, \mn@doi [Research in Astronomy and Astrophysics] {10.1088/1674-4527/acd89e}, \href {https://ui.adsabs.harvard.edu/abs/2023RAA....23h2001L} {23, 082001}

\bibitem[\protect\citeauthoryear{{Livio} \& {Mazzali}}{{Livio} \& {Mazzali}}{2018}]{2018PhR...736....1L}
{Livio} M.,  {Mazzali} P.,  2018, \mn@doi [\physrep] {10.1016/j.physrep.2018.02.002}, \href {https://ui.adsabs.harvard.edu/abs/2018PhR...736....1L} {736, 1}

\bibitem[\protect\citeauthoryear{{Maeda}, {Jiang}, {Shigeyama}  \& {Doi}}{{Maeda} et~al.}{2018}]{2018ApJ...861...78M}
{Maeda} K.,  {Jiang} J.-a.,  {Shigeyama} T.,   {Doi} M.,  2018, \mn@doi [\apj] {10.3847/1538-4357/aac8d8}, \href {https://ui.adsabs.harvard.edu/abs/2018ApJ...861...78M} {861, 78}

\bibitem[\protect\citeauthoryear{{Magee} et~al.,}{{Magee} et~al.}{2022}]{2022MNRAS.513.3035M}
{Magee} M.~R.,  et~al., 2022, \mn@doi [\mnras] {10.1093/mnras/stac1045}, \href {https://ui.adsabs.harvard.edu/abs/2022MNRAS.513.3035M} {513, 3035}

\bibitem[\protect\citeauthoryear{{Maguire} et~al.,}{{Maguire} et~al.}{2011}]{2011MNRAS.418..747M}
{Maguire} K.,  et~al., 2011, \mn@doi [\mnras] {10.1111/j.1365-2966.2011.19526.x}, \href {https://ui.adsabs.harvard.edu/abs/2011MNRAS.418..747M} {418, 747}

\bibitem[\protect\citeauthoryear{{Maguire}, {Chu}, {Dahiwale}  \& {Fremling}}{{Maguire} et~al.}{2022}]{2022TNSCR2845....1M}
{Maguire} K.,  {Chu} M.,  {Dahiwale} A.,   {Fremling} C.,  2022, Transient Name Server Classification Report, \href {https://ui.adsabs.harvard.edu/abs/2022TNSCR2845....1M} {2022-2845, 1}

\bibitem[\protect\citeauthoryear{{Marquardt}, {Sim}, {Ruiter}, {Seitenzahl}, {Ohlmann}, {Kromer}, {Pakmor}  \& {R{\"o}pke}}{{Marquardt} et~al.}{2015}]{2015A&A...580A.118M}
{Marquardt} K.~S.,  {Sim} S.~A.,  {Ruiter} A.~J.,  {Seitenzahl} I.~R.,  {Ohlmann} S.~T.,  {Kromer} M.,  {Pakmor} R.,   {R{\"o}pke} F.~K.,  2015, \mn@doi [\aap] {10.1051/0004-6361/201525761}, \href {https://ui.adsabs.harvard.edu/abs/2015A&A...580A.118M} {580, A118}

\bibitem[\protect\citeauthoryear{{Masci} et~al.,}{{Masci} et~al.}{2019}]{2019PASP..131a8003M}
{Masci} F.~J.,  et~al., 2019, \mn@doi [\pasp] {10.1088/1538-3873/aae8ac}, \href {https://ui.adsabs.harvard.edu/abs/2019PASP..131a8003M} {131, 018003}

\bibitem[\protect\citeauthoryear{{Mazzali}, {Chugai}, {Turatto}, {Lucy}, {Danziger}, {Cappellaro}, {della Valle}  \& {Benetti}}{{Mazzali} et~al.}{1997}]{1997MNRAS.284..151M}
{Mazzali} P.~A.,  {Chugai} N.,  {Turatto} M.,  {Lucy} L.~B.,  {Danziger} I.~J.,  {Cappellaro} E.,  {della Valle} M.,   {Benetti} S.,  1997, \mn@doi [\mnras] {10.1093/mnras/284.1.151}, \href {https://ui.adsabs.harvard.edu/abs/1997MNRAS.284..151M} {284, 151}

\bibitem[\protect\citeauthoryear{Miller \& Stone}{Miller \& Stone}{1994}]{miller1994kast}
Miller J.,  Stone R.,  1994, The Kast Double Spectograph.
Lick Observatory technical reports, University of California Observatories/Lick Observatory, \url {https://mthamilton.ucolick.org/techdocs/instruments/kast/Tech%20Report%2066%20KAST%20Miller%20Stone.pdf}

\bibitem[\protect\citeauthoryear{{Miller} et~al.,}{{Miller} et~al.}{2020}]{2020ApJ...898...56M}
{Miller} A.~A.,  et~al., 2020, \mn@doi [\apj] {10.3847/1538-4357/ab9e05}, \href {https://ui.adsabs.harvard.edu/abs/2020ApJ...898...56M} {898, 56}

\bibitem[\protect\citeauthoryear{{Miyaji}, {Nomoto}, {Yokoi}  \& {Sugimoto}}{{Miyaji} et~al.}{1980}]{1980PASJ...32..303M}
{Miyaji} S.,  {Nomoto} K.,  {Yokoi} K.,   {Sugimoto} D.,  1980, \pasj, \href {https://ui.adsabs.harvard.edu/abs/1980PASJ...32..303M} {32, 303}

\bibitem[\protect\citeauthoryear{{Nicholl}, {Guillochon}  \& {Berger}}{{Nicholl} et~al.}{2017}]{2017ApJ...850...55N}
{Nicholl} M.,  {Guillochon} J.,   {Berger} E.,  2017, \mn@doi [\apj] {10.3847/1538-4357/aa9334}, \href {https://ui.adsabs.harvard.edu/abs/2017ApJ...850...55N} {850, 55}

\bibitem[\protect\citeauthoryear{{Nugent}, {Kim}  \& {Perlmutter}}{{Nugent} et~al.}{2002}]{2002PASP..114..803N}
{Nugent} P.,  {Kim} A.,   {Perlmutter} S.,  2002, \mn@doi [\pasp] {10.1086/341707}, \href {https://ui.adsabs.harvard.edu/abs/2002PASP..114..803N} {114, 803}

\bibitem[\protect\citeauthoryear{{Nugent} et~al.,}{{Nugent} et~al.}{2011}]{2011Natur.480..344N}
{Nugent} P.~E.,  et~al., 2011, \mn@doi [\nat] {10.1038/nature10644}, \href {https://ui.adsabs.harvard.edu/abs/2011Natur.480..344N} {480, 344}

\bibitem[\protect\citeauthoryear{{Pakmor}, {Kromer}, {R{\"o}pke}, {Sim}, {Ruiter}  \& {Hillebrandt}}{{Pakmor} et~al.}{2010}]{2010Natur.463...61P}
{Pakmor} R.,  {Kromer} M.,  {R{\"o}pke} F.~K.,  {Sim} S.~A.,  {Ruiter} A.~J.,   {Hillebrandt} W.,  2010, \mn@doi [\nat] {10.1038/nature08642}, \href {https://ui.adsabs.harvard.edu/abs/2010Natur.463...61P} {463, 61}

\bibitem[\protect\citeauthoryear{{Parrent}, {Friesen}  \& {Parthasarathy}}{{Parrent} et~al.}{2014}]{2014Ap&SS.351....1P}
{Parrent} J.,  {Friesen} B.,   {Parthasarathy} M.,  2014, \mn@doi [\apss] {10.1007/s10509-014-1830-1}, \href {https://ui.adsabs.harvard.edu/abs/2014Ap&SS.351....1P} {351, 1}

\bibitem[\protect\citeauthoryear{{Pastorello} et~al.,}{{Pastorello} et~al.}{2007}]{2007MNRAS.377.1531P}
{Pastorello} A.,  et~al., 2007, \mn@doi [\mnras] {10.1111/j.1365-2966.2007.11700.x}, \href {https://ui.adsabs.harvard.edu/abs/2007MNRAS.377.1531P} {377, 1531}

\bibitem[\protect\citeauthoryear{{Pereira} et~al.,}{{Pereira} et~al.}{2013}]{2013A&A...554A..27P}
{Pereira} R.,  et~al., 2013, \mn@doi [\aap] {10.1051/0004-6361/201221008}, \href {https://ui.adsabs.harvard.edu/abs/2013A&A...554A..27P} {554, A27}

\bibitem[\protect\citeauthoryear{{Perlmutter} et~al.,}{{Perlmutter} et~al.}{1999}]{1999ApJ...517..565P}
{Perlmutter} S.,  et~al., 1999, \mn@doi [\apj] {10.1086/307221}, \href {https://ui.adsabs.harvard.edu/abs/1999ApJ...517..565P} {517, 565}

\bibitem[\protect\citeauthoryear{{Phillips}}{{Phillips}}{1993}]{1993ApJ...413L.105P}
{Phillips} M.~M.,  1993, \mn@doi [\apjl] {10.1086/186970}, \href {https://ui.adsabs.harvard.edu/abs/1993ApJ...413L.105P} {413, L105}

\bibitem[\protect\citeauthoryear{{Phillips}, {Lira}, {Suntzeff}, {Schommer}, {Hamuy}  \& {Maza}}{{Phillips} et~al.}{1999}]{1999AJ....118.1766P}
{Phillips} M.~M.,  {Lira} P.,  {Suntzeff} N.~B.,  {Schommer} R.~A.,  {Hamuy} M.,   {Maza} J.,  1999, \mn@doi [\aj] {10.1086/301032}, \href {https://ui.adsabs.harvard.edu/abs/1999AJ....118.1766P} {118, 1766}

\bibitem[\protect\citeauthoryear{{Piro} \& {Morozova}}{{Piro} \& {Morozova}}{2016}]{2016ApJ...826...96P}
{Piro} A.~L.,  {Morozova} V.~S.,  2016, \mn@doi [\apj] {10.3847/0004-637X/826/1/96}, \href {https://ui.adsabs.harvard.edu/abs/2016ApJ...826...96P} {826, 96}

\bibitem[\protect\citeauthoryear{{Piro} \& {Nakar}}{{Piro} \& {Nakar}}{2013}]{2013ApJ...769...67P}
{Piro} A.~L.,  {Nakar} E.,  2013, \mn@doi [\apj] {10.1088/0004-637X/769/1/67}, \href {https://ui.adsabs.harvard.edu/abs/2013ApJ...769...67P} {769, 67}

\bibitem[\protect\citeauthoryear{{Piro}, {Haynie}  \& {Yao}}{{Piro} et~al.}{2021}]{2021ApJ...909..209P}
{Piro} A.~L.,  {Haynie} A.,   {Yao} Y.,  2021, \mn@doi [\apj] {10.3847/1538-4357/abe2b1}, \href {https://ui.adsabs.harvard.edu/abs/2021ApJ...909..209P} {909, 209}

\bibitem[\protect\citeauthoryear{{Polin}, {Nugent}  \& {Kasen}}{{Polin} et~al.}{2019}]{2019ApJ...873...84P}
{Polin} A.,  {Nugent} P.,   {Kasen} D.,  2019, \mn@doi [\apj] {10.3847/1538-4357/aafb6a}, \href {https://ui.adsabs.harvard.edu/abs/2019ApJ...873...84P} {873, 84}

\bibitem[\protect\citeauthoryear{{Raskin} \& {Kasen}}{{Raskin} \& {Kasen}}{2013}]{2013ApJ...772....1R}
{Raskin} C.,  {Kasen} D.,  2013, \mn@doi [\apj] {10.1088/0004-637X/772/1/1}, \href {https://ui.adsabs.harvard.edu/abs/2013ApJ...772....1R} {772, 1}

\bibitem[\protect\citeauthoryear{{Riess} et~al.,}{{Riess} et~al.}{1998}]{1998AJ....116.1009R}
{Riess} A.~G.,  et~al., 1998, \mn@doi [\aj] {10.1086/300499}, \href {https://ui.adsabs.harvard.edu/abs/1998AJ....116.1009R} {116, 1009}

\bibitem[\protect\citeauthoryear{{Riess} et~al.,}{{Riess} et~al.}{1999}]{1999AJ....118.2675R}
{Riess} A.~G.,  et~al., 1999, \mn@doi [\aj] {10.1086/301143}, \href {https://ui.adsabs.harvard.edu/abs/1999AJ....118.2675R} {118, 2675}

\bibitem[\protect\citeauthoryear{{Riess} et~al.,}{{Riess} et~al.}{2022}]{2022ApJ...934L...7R}
{Riess} A.~G.,  et~al., 2022, \mn@doi [\apjl] {10.3847/2041-8213/ac5c5b}, \href {https://ui.adsabs.harvard.edu/abs/2022ApJ...934L...7R} {934, L7}

\bibitem[\protect\citeauthoryear{{Sai} et~al.,}{{Sai} et~al.}{2022}]{2022MNRAS.514.3541S}
{Sai} H.,  et~al., 2022, \mn@doi [\mnras] {10.1093/mnras/stac1525}, \href {https://ui.adsabs.harvard.edu/abs/2022MNRAS.514.3541S} {514, 3541}

\bibitem[\protect\citeauthoryear{{Scalzo} et~al.,}{{Scalzo} et~al.}{2010}]{2010ApJ...713.1073S}
{Scalzo} R.~A.,  et~al., 2010, \mn@doi [\apj] {10.1088/0004-637X/713/2/1073}, \href {https://ui.adsabs.harvard.edu/abs/2010ApJ...713.1073S} {713, 1073}

\bibitem[\protect\citeauthoryear{{Schlafly} \& {Finkbeiner}}{{Schlafly} \& {Finkbeiner}}{2011}]{2011ApJ...737..103S}
{Schlafly} E.~F.,  {Finkbeiner} D.~P.,  2011, \mn@doi [\apj] {10.1088/0004-637X/737/2/103}, \href {https://ui.adsabs.harvard.edu/abs/2011ApJ...737..103S} {737, 103}

\bibitem[\protect\citeauthoryear{{Schmidt} et~al.,}{{Schmidt} et~al.}{1998}]{1998ApJ...507...46S}
{Schmidt} B.~P.,  et~al., 1998, \mn@doi [\apj] {10.1086/306308}, \href {https://ui.adsabs.harvard.edu/abs/1998ApJ...507...46S} {507, 46}

\bibitem[\protect\citeauthoryear{{Scolnic} et~al.,}{{Scolnic} et~al.}{2018}]{2018ApJ...859..101S}
{Scolnic} D.~M.,  et~al., 2018, \mn@doi [\apj] {10.3847/1538-4357/aab9bb}, \href {https://ui.adsabs.harvard.edu/abs/2018ApJ...859..101S} {859, 101}

\bibitem[\protect\citeauthoryear{{Shappee} et~al.,}{{Shappee} et~al.}{2019}]{2019ApJ...870...13S}
{Shappee} B.~J.,  et~al., 2019, \mn@doi [\apj] {10.3847/1538-4357/aaec79}, \href {https://ui.adsabs.harvard.edu/abs/2019ApJ...870...13S} {870, 13}

\bibitem[\protect\citeauthoryear{{Shen}, {Bildsten}, {Kasen}  \& {Quataert}}{{Shen} et~al.}{2012}]{2012ApJ...748...35S}
{Shen} K.~J.,  {Bildsten} L.,  {Kasen} D.,   {Quataert} E.,  2012, \mn@doi [\apj] {10.1088/0004-637X/748/1/35}, \href {https://ui.adsabs.harvard.edu/abs/2012ApJ...748...35S} {748, 35}

\bibitem[\protect\citeauthoryear{{Siebert}, {Dimitriadis}, {Polin}  \& {Foley}}{{Siebert} et~al.}{2020}]{2020ApJ...900L..27S}
{Siebert} M.~R.,  {Dimitriadis} G.,  {Polin} A.,   {Foley} R.~J.,  2020, \mn@doi [\apjl] {10.3847/2041-8213/abae6e}, \href {https://ui.adsabs.harvard.edu/abs/2020ApJ...900L..27S} {900, L27}

\bibitem[\protect\citeauthoryear{{Smith} et~al.,}{{Smith} et~al.}{2020}]{2020PASP..132h5002S}
{Smith} K.~W.,  et~al., 2020, \mn@doi [\pasp] {10.1088/1538-3873/ab936e}, \href {https://ui.adsabs.harvard.edu/abs/2020PASP..132h5002S} {132, 085002}

\bibitem[\protect\citeauthoryear{{Srivastav} et~al.,}{{Srivastav} et~al.}{2023}]{2023arXiv230806019S}
{Srivastav} S.,  et~al., 2023, \mn@doi [arXiv e-prints] {10.48550/arXiv.2308.06019}, \href {https://ui.adsabs.harvard.edu/abs/2023arXiv230806019S} {p. arXiv:2308.06019}

\bibitem[\protect\citeauthoryear{{Stritzinger} \& {Leibundgut}}{{Stritzinger} \& {Leibundgut}}{2005}]{2005A&A...431..423S}
{Stritzinger} M.,  {Leibundgut} B.,  2005, \mn@doi [\aap] {10.1051/0004-6361:20041630}, \href {https://ui.adsabs.harvard.edu/abs/2005A&A...431..423S} {431, 423}

\bibitem[\protect\citeauthoryear{{Taubenberger}}{{Taubenberger}}{2017}]{2017hsn..book..317T}
{Taubenberger} S.,  2017, in {Alsabti} A.~W.,  {Murdin} P.,  eds, , Handbook of Supernovae.
p.~317, \mn@doi{10.1007/978-3-319-21846-5\_37}

\bibitem[\protect\citeauthoryear{{Taubenberger}, {Kromer}, {Pakmor}, {Pignata}, {Maeda}, {Hachinger}, {Leibundgut}  \& {Hillebrandt}}{{Taubenberger} et~al.}{2013}]{2013ApJ...775L..43T}
{Taubenberger} S.,  {Kromer} M.,  {Pakmor} R.,  {Pignata} G.,  {Maeda} K.,  {Hachinger} S.,  {Leibundgut} B.,   {Hillebrandt} W.,  2013, \mn@doi [\apjl] {10.1088/2041-8205/775/2/L43}, \href {https://ui.adsabs.harvard.edu/abs/2013ApJ...775L..43T} {775, L43}

\bibitem[\protect\citeauthoryear{{Thomas}, {Nugent}  \& {Meza}}{{Thomas} et~al.}{2011}]{2011PASP..123..237T}
{Thomas} R.~C.,  {Nugent} P.~E.,   {Meza} J.~C.,  2011, \mn@doi [\pasp] {10.1086/658673}, \href {https://ui.adsabs.harvard.edu/abs/2011PASP..123..237T} {123, 237}

\bibitem[\protect\citeauthoryear{{Tonry} et~al.,}{{Tonry} et~al.}{2018}]{2018PASP..130f4505T}
{Tonry} J.~L.,  et~al., 2018, \mn@doi [\pasp] {10.1088/1538-3873/aabadf}, \href {https://ui.adsabs.harvard.edu/abs/2018PASP..130f4505T} {130, 064505}

\bibitem[\protect\citeauthoryear{{Tucker}}{{Tucker}}{2022}]{2022TNSCR2778....1T}
{Tucker} M.~A.,  2022, Transient Name Server Classification Report, \href {https://ui.adsabs.harvard.edu/abs/2022TNSCR2778....1T} {2022-2778, 1}

\bibitem[\protect\citeauthoryear{{Tucker} et~al.,}{{Tucker} et~al.}{2021}]{2021ApJ...914...50T}
{Tucker} M.~A.,  et~al., 2021, \mn@doi [\apj] {10.3847/1538-4357/abf93b}, \href {https://ui.adsabs.harvard.edu/abs/2021ApJ...914...50T} {914, 50}

\bibitem[\protect\citeauthoryear{{Tucker} et~al.,}{{Tucker} et~al.}{2022}]{2022PASP..134l4502T}
{Tucker} M.~A.,  et~al., 2022, \mn@doi [\pasp] {10.1088/1538-3873/aca719}, \href {https://ui.adsabs.harvard.edu/abs/2022PASP..134l4502T} {134, 124502}

\bibitem[\protect\citeauthoryear{{Turatto}, {Benetti}, {Cappellaro}, {Danziger}, {Della Valle}, {Gouiffes}, {Mazzali}  \& {Patat}}{{Turatto} et~al.}{1996}]{1996MNRAS.283....1T}
{Turatto} M.,  {Benetti} S.,  {Cappellaro} E.,  {Danziger} I.~J.,  {Della Valle} M.,  {Gouiffes} C.,  {Mazzali} P.~A.,   {Patat} F.,  1996, \mn@doi [\mnras] {10.1093/mnras/283.1.1}, \href {https://ui.adsabs.harvard.edu/abs/1996MNRAS.283....1T} {283, 1}

\bibitem[\protect\citeauthoryear{{Wang}, {Wang}, {Zhou}, {Lou}  \& {Li}}{{Wang} et~al.}{2005}]{2005ApJ...620L..87W}
{Wang} X.,  {Wang} L.,  {Zhou} X.,  {Lou} Y.-Q.,   {Li} Z.,  2005, \mn@doi [\apjl] {10.1086/428774}, \href {https://ui.adsabs.harvard.edu/abs/2005ApJ...620L..87W} {620, L87}

\bibitem[\protect\citeauthoryear{{Wang} et~al.,}{{Wang} et~al.}{2008}]{2008ApJ...675..626W}
{Wang} X.,  et~al., 2008, \mn@doi [\apj] {10.1086/526413}, \href {https://ui.adsabs.harvard.edu/abs/2008ApJ...675..626W} {675, 626}

\bibitem[\protect\citeauthoryear{{Wang} et~al.,}{{Wang} et~al.}{2009}]{2009ApJ...699L.139W}
{Wang} X.,  et~al., 2009, \mn@doi [\apjl] {10.1088/0004-637X/699/2/L139}, \href {https://ui.adsabs.harvard.edu/abs/2009ApJ...699L.139W} {699, L139}

\bibitem[\protect\citeauthoryear{{Wang}, {Wang}, {Filippenko}, {Zhang}  \& {Zhao}}{{Wang} et~al.}{2013}]{2013Sci...340..170W}
{Wang} X.,  {Wang} L.,  {Filippenko} A.~V.,  {Zhang} T.,   {Zhao} X.,  2013, \mn@doi [Science] {10.1126/science.1231502}, \href {https://ui.adsabs.harvard.edu/abs/2013Sci...340..170W} {340, 170}

\bibitem[\protect\citeauthoryear{{Wang}, {Chen}, {Wang}, {Hu}, {Xi}, {Yang}, {Zhao}  \& {Li}}{{Wang} et~al.}{2019}]{2019ApJ...882..120W}
{Wang} X.,  {Chen} J.,  {Wang} L.,  {Hu} M.,  {Xi} G.,  {Yang} Y.,  {Zhao} X.,   {Li} W.,  2019, \mn@doi [\apj] {10.3847/1538-4357/ab26b5}, \href {https://ui.adsabs.harvard.edu/abs/2019ApJ...882..120W} {882, 120}

\bibitem[\protect\citeauthoryear{{Wang} et~al.,}{{Wang} et~al.}{2023}]{2023arXiv230503779W}
{Wang} Q.,  et~al., 2023, \mn@doi [arXiv e-prints] {10.48550/arXiv.2305.03779}, \href {https://ui.adsabs.harvard.edu/abs/2023arXiv230503779W} {p. arXiv:2305.03779}

\bibitem[\protect\citeauthoryear{{Webbink}}{{Webbink}}{1984}]{1984ApJ...277..355W}
{Webbink} R.~F.,  1984, \mn@doi [\apj] {10.1086/161701}, \href {https://ui.adsabs.harvard.edu/abs/1984ApJ...277..355W} {277, 355}

\bibitem[\protect\citeauthoryear{{Whelan} \& {Iben}}{{Whelan} \& {Iben}}{1973}]{1973ApJ...186.1007W}
{Whelan} J.,  {Iben} Icko J.,  1973, \mn@doi [\apj] {10.1086/152565}, \href {https://ui.adsabs.harvard.edu/abs/1973ApJ...186.1007W} {186, 1007}

\bibitem[\protect\citeauthoryear{{White} et~al.,}{{White} et~al.}{2015}]{2015ApJ...799...52W}
{White} C.~J.,  et~al., 2015, \mn@doi [\apj] {10.1088/0004-637X/799/1/52}, \href {https://ui.adsabs.harvard.edu/abs/2015ApJ...799...52W} {799, 52}

\bibitem[\protect\citeauthoryear{{Wilkinson}, {Maraston}, {Goddard}, {Thomas}  \& {Parikh}}{{Wilkinson} et~al.}{2017}]{2017MNRAS.472.4297W}
{Wilkinson} D.~M.,  {Maraston} C.,  {Goddard} D.,  {Thomas} D.,   {Parikh} T.,  2017, \mn@doi [\mnras] {10.1093/mnras/stx2215}, \href {https://ui.adsabs.harvard.edu/abs/2017MNRAS.472.4297W} {472, 4297}

\bibitem[\protect\citeauthoryear{{Willcox}, {Townsley}, {Calder}, {Denissenkov}  \& {Herwig}}{{Willcox} et~al.}{2016}]{2016ApJ...832...13W}
{Willcox} D.~E.,  {Townsley} D.~M.,  {Calder} A.~C.,  {Denissenkov} P.~A.,   {Herwig} F.,  2016, \mn@doi [\apj] {10.3847/0004-637X/832/1/13}, \href {https://ui.adsabs.harvard.edu/abs/2016ApJ...832...13W} {832, 13}

\bibitem[\protect\citeauthoryear{{Xi} et~al.,}{{Xi} et~al.}{2022}]{2022MNRAS.517.4098X}
{Xi} G.,  et~al., 2022, \mn@doi [\mnras] {10.1093/mnras/stac2848}, \href {https://ui.adsabs.harvard.edu/abs/2022MNRAS.517.4098X} {517, 4098}

\bibitem[\protect\citeauthoryear{{Yoon}, {Podsiadlowski}  \& {Rosswog}}{{Yoon} et~al.}{2007}]{2007MNRAS.380..933Y}
{Yoon} S.~C.,  {Podsiadlowski} P.,   {Rosswog} S.,  2007, \mn@doi [\mnras] {10.1111/j.1365-2966.2007.12161.x}, \href {https://ui.adsabs.harvard.edu/abs/2007MNRAS.380..933Y} {380, 933}

\bibitem[\protect\citeauthoryear{{York} et~al.,}{{York} et~al.}{2000}]{2000AJ....120.1579Y}
{York} D.~G.,  et~al., 2000, \mn@doi [\aj] {10.1086/301513}, \href {https://ui.adsabs.harvard.edu/abs/2000AJ....120.1579Y} {120, 1579}

\bibitem[\protect\citeauthoryear{{Zhang}, {Wang}, {Zhou}, {Li}, {Ma}, {Jiang}  \& {Li}}{{Zhang} et~al.}{2004}]{2004AJ....128.1857Z}
{Zhang} T.,  {Wang} X.,  {Zhou} X.,  {Li} W.,  {Ma} J.,  {Jiang} Z.,   {Li} Z.,  2004, \mn@doi [\aj] {10.1086/423699}, \href {https://ui.adsabs.harvard.edu/abs/2004AJ....128.1857Z} {128, 1857}

\bibitem[\protect\citeauthoryear{{Zhang} et~al.,}{{Zhang} et~al.}{2016}]{2016ApJ...820...67Z}
{Zhang} K.,  et~al., 2016, \mn@doi [\apj] {10.3847/0004-637X/820/1/67}, \href {https://ui.adsabs.harvard.edu/abs/2016ApJ...820...67Z} {820, 67}

\bibitem[\protect\citeauthoryear{{Zhao} et~al.,}{{Zhao} et~al.}{2016}]{2016ApJ...826..211Z}
{Zhao} X.,  et~al., 2016, \mn@doi [\apj] {10.3847/0004-637X/826/2/211}, \href {https://ui.adsabs.harvard.edu/abs/2016ApJ...826..211Z} {826, 211}

\bibitem[\protect\citeauthoryear{{van Kerkwijk}, {Chang}  \& {Justham}}{{van Kerkwijk} et~al.}{2010}]{2010ApJ...722L.157V}
{van Kerkwijk} M.~H.,  {Chang} P.,   {Justham} S.,  2010, \mn@doi [\apjl] {10.1088/2041-8205/722/2/L157}, \href {https://ui.adsabs.harvard.edu/abs/2010ApJ...722L.157V} {722, L157}

\makeatother
\end{thebibliography}

\begin{table*}
\caption{Spectroscopic Observations of SN~2022vqz\label{tabspec}}
\begin{threeparttable}[b]
\begin{tabular}{c c c c c c c c}
\hline
   UT  & MJD & Epoch\tnote{a} & Telescope & Instrument & Range(\AA) & Resolution(\AA)\\
\hline
   2022-10-04 & 59856.25 & $-7.1$ & Lick & Kast & 3630--10,760 & 2.0\\
   2022-10-04 & 59856.54 & $-6.8$ & XLT & BFOSC & 3900--8870 & 2.8\\
   2022-10-07 & 59859.53 & $-3.8$ & XLT & BFOSC & 3900--8870 & 2.8\\
   2022-10-09 & 59861.73 & $-1.7$ & XLT & BFOSC & 3910--8870 & 2.8\\
   2022-10-10 & 59862.70 & $-0.7$ & XLT & BFOSC & 3910--8870 & 2.8\\
   2022-10-14 & 59866.63 & $3.1$ & XLT & BFOSC & 3910--8880 & 2.8\\
   2022-10-17 & 59869.70 & $6.2$ & XLT & BFOSC & 3910--8870 & 2.8\\
   2022-10-18 & 59870.63 & $7.1$ & XLT & BFOSC & 3900--8860 & 2.8\\
   2022-10-19 & 59871.75 & $8.2$ & Copernico & AFOSC & 3560--9290 & 4.0\\
   2022-10-21 & 59873.65 & $10.0$ & XLT & BFOSC & 3900--8870 & 2.8\\
   2022-10-27 & 59879.33 & $15.6$ & Lick & Kast & 3640--10,750 & 2.0\\
   2022-11-04 & 59887.54 & $23.7$ & XLT & BFOSC & 3910--8870 & 2.8\\
   2022-11-10 & 59893.69 & $29.8$ & LJT & YFOSC & 3530--8780 & 2.8\\
   2022-11-17 & 59900.18 & $36.1$ & Lick & Kast & 3640--10,740 & 2.0\\
   2022-11-22 & 59905.67 & $41.5$ & LJT & YFOSC & 3520--8780 & 2.8\\
   2022-12-08 & 59921.57 & $57.2$ & LJT & YFOSC & 3530--8790 & 2.8\\
   2022-12-10 & 59923.54 & $59.1$\tnote{b} & XLT & BFOSC & 3910--8870 & 2.8\\
   2023-01-10 & 59954.57 & $89.6$ & LJT & YFOSC & 3640--8940 & 2.9\\
   2023-01-18 & 59962.50 & $97.4$ & LJT & YFOSC & 3630--8940 & 2.9\\
\hline
\end{tabular}
\begin{tablenotes}
\item[a]{Days relative to $B$-band maximum brightness on 2022-10-11.43 (MJD~59863.43), corrected to the rest frame by the factor $1+z=1.017$.}
\item[b]{The spectrum of the host galaxy.}
\end{tablenotes}
\end{threeparttable}
\end{table*}

\begin{table*}
\caption{Strength and Weakness of Models for SN~2022vqz\label{tabmodels}}
\begin{threeparttable}[b]
\begin{tabular}{c c c c c c}
\hline
                         &    Low     &   Low    &    Early   &   $g-r$    &   Carbon   \\
                         & Luminosity & Velocity &    Excess  & Evolution  & Deficiency \\
\hline
Double Detonation        &            &          & \checkmark & \checkmark &            \\
+ Hybrid CONe WD         &     ?      &\checkmark&            &            & \checkmark \\
\hline
Violent Merger           & \checkmark &          &            &            &            \\
+ CSM Interaction        &            &          & \checkmark & \checkmark &  $\times$  \\
\hline
Companion Interaction    &            &          & \checkmark & $\times$   &            \\
\hline
Ni Mixing                &            &          &  $\times$  & $\times$   &            \\
\hline
\end{tabular}
\begin{tablenotes}
\item[]{\checkmark means strength and $\times$ means weakness. A question mark denotes an unresolved tension. Detailed reasonings are described in the text. }
\end{tablenotes}
\end{threeparttable}
\end{table*}

\begin{table*}
\caption{SN~2022vqz Photometry from TNT\label{tablctnt}}
\begin{threeparttable}[b]
\begin{tabular}{c c c c c c}
\hline
   MJD & $B$ (mag) & $V$ (mag) & $g$ (mag) & $r$ (mag) & $i$ (mag) \\
\hline
59849.80 & 18.813(211) & 18.462(165) & 18.688(161) & 18.479(134) & 18.346(119)\\
59856.81 & 16.864(183) & 16.485(156) & 16.617(144) & 16.495(124) & 16.584(113)\\
59858.83 & 16.568(195) & 16.204(170) & 16.313(160) & 16.233(132) & 16.351(113)\\
59861.74 & 16.262(162) & 15.911(115) & 16.015(116) & 15.777(102) & 15.982(094)\\
59862.73 & 16.221(180) & 15.887(145) & 15.976(132) & 15.917(116) & 16.018(134)\\
59863.72 & 16.234(172) & 15.853(126) & 15.975(136) & 15.839(131) & 16.078(144)\\
59866.75 & 16.322(174) & 15.852(142) & 16.011(160) & 15.868(146) & 16.123(116)\\
59868.73 & 16.453(161) & 15.896(122) & 16.137(112) & 15.835(100) & 16.101(105)\\
59869.82 & 16.593(168) & 15.950(130) & 16.204(127) & 15.860(117) & -\\
59873.61 & 16.945(195) & 16.206(145) & 16.598(160) & 16.073(132) & 16.271(116)\\
59874.67 & 17.106(188) & 16.274(155) & 16.696(160) & 16.112(131) & 16.328(113)\\
59875.70 & 17.242(184) & 16.336(146) & 16.840(150) & 16.148(129) & 16.353(121)\\
59876.68 & 17.362(188) & 16.432(147) & 16.949(137) & 16.234(121) & 16.344(120)\\
59877.65 & 17.428(172) & 16.436(139) & 17.039(147) & 16.284(130) & 16.393(126)\\
59883.72 & 18.203(165) & 16.920(117) & 17.644(115) & 16.500(094) & 16.538(099)\\
59884.49 & 17.959(204) & 17.066(167) & - & - & -\\
59886.49 & 17.988(171) & 17.033(128) & 17.823(122) & 16.558(116) & 16.477(115)\\
59895.71 & 18.464(171) & 17.314(112) & - & - & -\\
59897.65 & 18.694(204) & 17.657(163) & 18.441(158) & 17.363(132) & 17.247(116)\\
59898.67 & 18.578(175) & 17.603(137) & 18.138(128) & 17.121(111) & 17.002(111)\\
59902.59 & 18.894(212) & 17.705(164) & 18.468(161) & 17.526(133) & 17.413(116)\\
59906.45 & 18.969(196) & 17.892(163) & 18.577(155) & 17.686(133) & 17.476(132)\\
59908.56 & 18.750(186) & 18.078(134) & 18.719(133) & 17.672(119) & 17.559(117)\\
\hline
\end{tabular}
\begin{tablenotes}
\item[] {Uncertainties, in units of 0.001~mag, are 1$\sigma$.}
\end{tablenotes}
\end{threeparttable}
\end{table*}

\clearpage

\begin{table*}
\caption{SN~2022vqz Photometry from AZT\label{tablcazt}}
\begin{threeparttable}[b]
\begin{tabular}{c c c c c c}
\hline
   MJD & $U$ (mag) & $B$ (mag) & $V$ (mag) & $R$ (mag) & $I$ (mag) \\
\hline
59853.80 & 17.145(137) & 17.340(044) & 16.939(033) & 16.667(024) & 16.686(025)\\
59855.73 & 16.759(130) & 16.897(040) & 16.561(031) & 16.383(022) & 16.280(024)\\
59856.75 & 16.453(162) & 16.770(053) & 16.465(040) & 16.228(030) & 16.131(029)\\
59858.91 & 16.134(128) & 16.475(041) & 16.175(031) & 15.977(021) & 15.933(024)\\
59859.93 & 16.069(144) & 16.357(046) & 16.045(035) & 15.858(026) & 15.825(026)\\
59860.80 & 16.056(144) & 16.289(046) & 15.999(035) & 15.804(026) & 15.760(026)\\
59861.95 & 15.947(126) & 16.183(040) & 15.904(030) & 15.725(022) & 15.701(024)\\
59862.88 & 15.944(126) & 16.192(046) & 15.862(031) & 15.684(022) & 15.665(024)\\
59866.88 & 16.149(144) & 16.265(046) & 15.815(035) & 15.613(026) & 15.628(026)\\
59868.91 & 16.433(162) & 16.271(053) & 15.882(040) & 15.693(030) & 15.642(029)\\
59872.82 & 16.982(162) & 16.843(053) & 16.091(040) & 15.788(030) & 15.750(029)\\
59873.72 & 17.123(162) & 16.968(053) & 16.159(040) & 15.832(030) & 15.775(029)\\
59892.82 & 18.016(148) & 18.672(047) & 17.431(036) & 16.789(026) & 16.409(025)\\
59909.66 & - & 18.870(057) & 17.792(072) & 17.409(061) & -\\
59910.65 & 19.321(174) & 19.122(054) & 17.985(041) & 17.521(031) & 17.063(029)\\
59916.71 & - & 19.605(099) & 18.393(046) & 17.904(029) & 17.468(029)\\
59923.71 & - & 19.504(054) & 18.460(034) & 18.113(025) & 17.717(027)\\
\hline
\end{tabular}
\begin{tablenotes}
\item[] {Uncertainties, in units of 0.001~mag, are 1$\sigma$.}
\end{tablenotes}
\end{threeparttable}
\end{table*}

\begin{table*}
\caption{SN~2022vqz Photometry from Padova\label{tablcpadova}}
\begin{threeparttable}[b]
\begin{tabular}{c c c c c c c}
\hline
   MJD & Telescope\tnote{a} & $u$ (mag) & $g$ (mag) & $r$ (mag) & $i$ (mag) & $z$ (mag)\\
\hline
59859.03 & S & 16.810(198) & 16.261(160) & 16.112(151) & 16.299(155) & -\\
59864.98 & S & - & - & - & 15.986(152) & -\\
59869.03 & S & 17.246(203) & 16.144(157) & 15.842(121) & - & -\\
59869.79 & C & - & - & - & 16.146(113) & 16.174(163)\\
59874.96 & S & 18.127(209) & 16.648(148) & 16.054(163) & 16.263(150) & -\\
59880.95 & S & 19.018(211) & 17.334(157) & 16.369(121) & 16.438(121) & -\\
59888.85 & S & - & 17.781(150) & 16.755(149) & 16.656(145) & -\\
59893.98 & S & - & 18.325(147) & 16.979(149) & 16.933(139) & -\\
59901.72 & C & - & 18.456(159) & 17.383(138) & 17.306(113) & 17.059(164)\\
59909.93 & S & - & 18.654(134) & 17.852(116) & 17.703(127) & -\\
59925.92 & C & - & - & - & 17.668(126) & -\\
\hline
\end{tabular}
\begin{tablenotes}
\item[a]{S: Schmidt 67/92~cm Telescope; C: 1.82~m Copernico Telescope.}
\item[] {Uncertainties, in units of 0.001~mag, are 1$\sigma$.}
\end{tablenotes}
\end{threeparttable}
\end{table*}

\clearpage

\begin{table*}
\caption{SN~2022vqz Photometry from ZTF\label{tablcztf}}
\begin{threeparttable}[b]
\begin{tabular}{c c c |c c c |c c c}
\hline
   MJD & filter & mag & MJD & filter & mag & MJD & filter & mag\\
\hline
59846.369 & $g$ & 18.379(074) & 59866.359 & $r$ & 15.731(032) & 59902.279 & $g$ & 18.247(097)\\
59846.425 & $r$ & 18.493(067) & 59870.297 & $r$ & 15.808(031) & 59904.175 & $g$ & 18.288(104)\\
59848.346 & $r$ & 18.898(085) & 59870.354 & $g$ & 16.133(035) & 59904.193 & $r$ & 17.372(050)\\
59848.402 & $g$ & 19.187(112) & 59872.321 & $g$ & 16.305(042) & 59906.152 & $r$ & 17.451(060)\\
59850.275 & $r$ & 18.001(063) & 59872.366 & $r$ & 15.891(031) & 59906.190 & $g$ & 18.339(071)\\
59850.388 & $g$ & 18.080(082) & 59874.216 & $g$ & 16.536(045) & 59909.195 & $g$ & 18.400(086)\\
59852.320 & $r$ & 17.284(042) & 59874.274 & $r$ & 15.979(029) & 59909.247 & $r$ & 17.550(066)\\
59852.348 & $g$ & 17.406(048) & 59877.351 & $r$ & 16.133(028) & 59911.190 & $r$ & 17.521(161)\\
59854.288 & $r$ & 16.817(052) & 59877.409 & $g$ & 16.867(048) & 59914.152 & $r$ & 17.732(085)\\
59854.385 & $g$ & 16.953(039) & 59880.220 & $r$ & 16.280(033) & 59914.219 & $g$ & 18.401(085)\\
59856.333 & $g$ & 16.556(039) & 59880.237 & $g$ & 17.177(047) & 59923.218 & $g$ & 18.603(113)\\
59856.348 & $r$ & 16.409(049) & 59882.223 & $g$ & 17.374(055) & 59928.152 & $r$ & 18.197(114)\\
59858.320 & $r$ & 16.123(036) & 59888.245 & $g$ & 17.713(078) & 59928.198 & $g$ & 18.747(129)\\
59858.407 & $g$ & 16.204(048) & 59888.319 & $r$ & 16.721(061) & 59932.196 & $g$ & 18.833(087)\\
59860.221 & $r$ & 15.955(030) & 59894.303 & $r$ & 16.942(043) & 59932.225 & $r$ & 18.302(100)\\
59860.340 & $g$ & 16.023(045) & 59896.221 & $g$ & 18.031(073) & 59934.239 & $r$ & 18.350(111)\\
59862.384 & $g$ & 15.894(047) & 59896.283 & $r$ & 17.036(057) & 59936.174 & $g$ & 18.769(132)\\
59864.324 & $r$ & 15.736(033) & 59898.228 & $g$ & 18.082(092) & 59936.199 & $r$ & 18.409(083)\\
59864.387 & $g$ & 15.850(040) & 59900.216 & $r$ & 17.228(073) &  &  & \\
59866.269 & $g$ & 15.886(037) & 59902.165 & $r$ & 17.316(046) &  &  & \\
\hline
\end{tabular}
\begin{tablenotes}
\item[] {Uncertainties, in units of 0.001~mag, are 1$\sigma$.}
\end{tablenotes}
\end{threeparttable}
\end{table*}

\begin{table*}
\caption{SN~2022vqz Photometry from ATLAS\label{tablcatlas}}
\begin{threeparttable}[b]
\begin{tabular}{c c c |c c c |c c c}
\hline
   MJD & filter & mag & MJD & filter & mag & MJD & filter & mag\\
\hline
59845.493 & $o$ & 19.086(084) & 59869.349 & $c$ & 15.954(006) & 59899.368 & $c$ & 17.678(019)\\
59846.418 & $o$ & 18.439(048) & 59870.293 & $c$ & 15.974(006) & 59900.377 & $c$ & 17.673(020)\\
59847.428 & $c$ & 19.165(054) & 59871.368 & $o$ & 15.958(006) & 59901.467 & $o$ & 17.273(019)\\
59848.505 & $c$ & 19.106(066) & 59872.424 & $o$ & 15.997(007) & 59902.294 & $o$ & 17.316(017)\\
59849.416 & $o$ & 18.317(039) & 59873.560 & $o$ & 16.004(007) & 59903.360 & $c$ & 17.736(028)\\
59850.525 & $o$ & 18.051(035) & 59875.405 & $o$ & 16.113(007) & 59904.411 & $c$ & 17.933(024)\\
59851.456 & $c$ & 17.697(018) & 59876.391 & $o$ & 16.164(007) & 59906.357 & $o$ & 17.413(021)\\
59852.485 & $c$ & 17.429(014) & 59879.407 & $o$ & 16.304(020) & 59907.313 & $c$ & 17.956(022)\\
59853.459 & $o$ & 17.027(014) & 59880.424 & $o$ & 16.270(020) & 59908.317 & $c$ & 18.015(024)\\
59854.389 & $o$ & 16.845(012) & 59881.402 & $c$ & 16.816(011) & 59909.359 & $o$ & 17.465(020)\\
59855.402 & $o$ & 16.660(010) & 59882.407 & $c$ & 16.859(012) & 59910.301 & $o$ & 17.759(144)\\
59856.457 & $o$ & 16.451(008) & 59883.387 & $o$ & 16.409(008) & 59914.351 & $o$ & 17.632(023)\\
59858.478 & $o$ & 16.175(010) & 59884.364 & $o$ & 16.435(009) & 59926.341 & $o$ & 18.071(043)\\
59859.437 & $o$ & 16.082(009) & 59885.389 & $o$ & 16.500(011) & 59957.348 & $o$ & 18.849(092)\\
59860.449 & $o$ & 15.996(014) & 59886.406 & $o$ & 16.521(012) & 59958.225 & $o$ & 18.641(094)\\
59864.522 & $o$ & 15.837(008) & 59893.483 & $o$ & 16.883(030) & 59965.257 & $o$ & 19.459(136)\\
59866.520 & $o$ & 15.796(007) & 59895.369 & $o$ & 16.920(014) & 59966.226 & $o$ & 19.269(117)\\
59867.473 & $o$ & 15.860(006) & 59896.370 & $o$ & 16.994(015) & 59970.310 & $o$ & 19.509(283)\\
59868.432 & $o$ & 15.857(006) & 59897.327 & $o$ & 17.058(010) &  &  & \\
\hline
\end{tabular}
\begin{tablenotes}
\item[] {Uncertainties, in units of 0.001~mag, are 1$\sigma$.}
\end{tablenotes}
\end{threeparttable}
\end{table*}

\clearpage

\clearpage

\clearpage

\clearpage

\clearpage

\clearpage

\clearpage

\clearpage

\clearpage

\clearpage

\clearpage

\clearpage

\clearpage

\clearpage

\bsp	
\label{lastpage}
\end{document}